\documentclass[journal, onecolumn, 11pt]{IEEEtran}
\renewcommand{\baselinestretch}{2.0}
\usepackage{cite}
\usepackage{ifpdf}
\usepackage{array}
\ifpdf
\usepackage{graphicx}
\usepackage[svgnames]{xcolor}
\else
\usepackage[dvipdfmx]{graphicx}
\usepackage[dvipdfmx,svgnames]{xcolor}
\fi
\usepackage{amsmath,amssymb,amscd}
\usepackage{verbatim}

\newtheorem{lemma}{Lemma}

\newtheorem{rem}{Remark}
\newtheorem{prop}{Proposition}
\usepackage{algorithm}
\usepackage{algpseudocode}

\begin{document}
\renewcommand{\baselinestretch}{1.55}
\title{Joint Signal and Channel State Information \\Compression for the Backhaul of Uplink \\Network MIMO Systems}

\author{\Large Jinkyu Kang, Osvaldo Simeone, Joonhyuk Kang and Shlomo Shamai (Shitz)
\thanks{Jinkyu Kang and Joonhyuk Kang are with the Department of Electrical Engineering, Korea Advanced Institute of Science and Technology (KAIST) Daejeon, South Korea (Email: kangjk@kaist.ac.kr and jhkang@ee.kaist.ac.kr).

O. Simeone is with the Center for Wireless Communications and Signal Processing Research (CWCSPR), ECE Department, New Jersey Institute of Technology (NJIT), Newark, NJ 07102, USA (Email: osvaldo.simeone@njit.edu). 

S. Shamai (Shitz) is with the Department of Electrical Engineering, Technion, Haifa, 32000, Israel (Email: sshlomo@ee.technion.ac.il).
}}
\maketitle
\renewcommand{\baselinestretch}{2.0}
\begin{abstract}
In network MIMO cellular systems, subsets of base stations (BSs), or remote radio heads, are connected via backhaul links to central units (CUs) that perform joint encoding in the downlink and joint decoding in the uplink. Focusing on the uplink, an effective solution for the communication between BSs and the corresponding CU on the backhaul links is based on compressing and forwarding the baseband received signal from each BS. In the presence of ergodic fading, communicating the channel state information (CSI) from the BSs to the CU may require a sizable part of the backhaul capacity. In a prior work, this aspect was studied by assuming a \textit{Compress-Forward-Estimate} (CFE) approach, whereby the BSs compress the training signal and CSI estimation takes place at the CU. In this work, instead, an \textit{Estimate-Compress-Forward} (ECF) approach is investigated, whereby the BSs perform CSI estimation and forward a compressed version of the CSI to the CU. This choice is motivated by the information theoretic optimality of separate estimation and compression. Various ECF strategies are proposed that perform either separate or joint compression of estimated CSI and received signal. Moreover, the proposed strategies are combined with distributed source coding when considering multiple BSs. ``Semi-coherent" strategies are also proposed that do not convey any CSI or training information on the backhaul links. Via numerical results, it is shown that a proper design of ECF strategies based on joint received signal and estimated CSI compression or of semi-coherent schemes leads to substantial performance gains compared to more conventional approaches based on non-coherent transmission or the CFE approach.
\end{abstract}


\section{Introduction}
In network MIMO systems, multiple base stations (BSs), or remote radio heads, are connected via backhaul links to a central unit (CU). Under ideal BSs-to-CU connectivity conditions, the CU performs joint encoding in downlink and joint decoding in uplink on behalf of all the connected BSs (see \cite{Gesbert10JSAC, SimeoneBookFnT, Marsch12VTMAG} and references therein). In the presence of practical limitations on the backhaul links, various strategies have been proposed for the communication between BSs and CU. Among these, one that appears to be favored due to its practicality and good theoretical performance is based on compress-and-forward \cite{LightradioAL, Sanderovich09TIT, Tian09TIT, Xie13TIT}. Accordingly, focusing on the uplink, the BSs compress the received baseband signal and forward it to the CU. Network MIMO with compress-and-forward BSs is also known as cloud radio access (see, e.g., \cite{IntelCor, LiuCOMMAG2011, ChinaMobile, FlanaganTI2011, Ericsson2012, HeathCOMMAG2011}).

Previous work on the design of backhaul compression strategies for the uplink has focused mostly on the problem of compressing the baseband received signal, and has implicitly assumed full channel state information (CSI) to be available at the CU \cite{Sanderovich09TIT, Coso09TWC, Park13TVT, Zhou12arXiv}. This assumption comes with little loss of generality in quasi-static channels in which the coherence time/bandwidth of the channel is large enough. In this case, in fact, the CSI overhead on the backhaul can be amortized within the channel coherence time. Instead, in the presence of time-varying or frequency selective channels, CSI overhead can become significant. Under this assumption, it is hence important to properly design the transfer of CSI and data from the BSs to the CU.

The backhaul overhead due to CSI transfer between BSs and CU in the uplink was studied in \cite{Kobayashi11TSP, Caire10ITA} by adapting the standard model of \cite{Hassibi03TIT}. Accordingly, the transmission period is divided into coherence intervals of limited lengths, each of which is used for both training and data transmission. It is recalled that, in \cite{Hassibi03TIT}, this model was used to study a point-to-point MIMO system, and then the analysis was extended for downlink MIMO systems (with no backhaul constraints) in \cite{Zheng02TIT, Kobayashi11TCOM}. Related work that concerns models in which BSs are connected to one another (see, e.g., \cite{Simeone09TIT, Aktas08TWC}) and CSI is imperfect can be found in \cite{Marsch09ICC, Marsch11TWC}.

In \cite{Kobayashi11TSP}, an uplink system is studied in which the received baseband signals are first compressed by each BS and then transmitted over the backhaul to the CU. The latter performs channel estimation based on the training part of the compressed received signals and then carries out joint decoding. We refer to this approach as \textit{Compress-Forward-Estimate} (CFE). In this work, we instead study an alternative approach that is motivated by the classical information-theoretic result concerning the separation of estimation and compression \cite{Witsenhausen80TIT}. This result states that, when compressing a noisy observation, it is optimal to first estimate the signal of interest and then compress the estimate, rather than to let the estimation be performed at the decoder's side. Following this insight, we propose various strategies that are based on an \textit{Estimate-Compress-Forward} (ECF) approach: each BS first estimates the CSI and then compresses it for transmission to the CU\footnote{The possibility to use an ECF approach rather than CFE was well recognized in \cite{Kobayashi11TSP}, where it is stated that: ``$\dots$ It is for example not clear if each BS should estimate its local channels and forward compressed versions of its estimates to the central station (CS) or if the CS should estimate all channels based on compressed signals from the BSs, $\dots$".}. Specifically, the proposed strategies carry out separate or joint compression of the estimated CSI and the received signal in the data part of the block. 

The main contributions in this paper are summarized as follows: 
\begin{itemize}
\item Proposal and analysis of a class of ECF strategies for the separate or joint compression of the estimated CSI and of the received data signal;
\item Proposal and analysis of a novel semi-coherent processing strategy that is based on the compression of the data signal after equalization at the BSs;
\item Thorough performance comparison among the non-coherent transmission scheme, the CFE method \cite{Kobayashi11TSP}, and the proposed ECF and semi-coherent strategies via numerical results.
\end{itemize}

The rest of the paper is organized as follows. We first review the conventional schemes, namely the non-coherent approach and the CFE scheme in Section \ref{Sec:Pre}. Then, we propose and analyze the ECF strategies in Section \ref{Sec:SBC} for the single-BS case and in Section \ref{Sec:MBC} for the more general scenario with multiple BSs. There, we combine the proposed ECF techniques with the distributed source coding strategies of \cite{Coso09TWC}. Moreover, in Section \ref{Sec:NCnSCP} we propose ``semi-coherent" schemes that do not convey any pilot information on the backhaul links. In Section \ref{Sec:Numerical Results}, numerical results are presented. Concluding remarks are summarized in Section \ref{Sec:Conclusion}.

\emph{Notation}: $E[ \cdot ]$, $\textrm{tr} (\cdot)$, and $\textrm{vec} (\cdot)$ denote the expectation,  trace, and vectorization (i.e., stacking of the columns) of the argument matrix. The Kronecker product is denoted by $\otimes$. We use the standard notation for mutual information and differential entropy \cite{GamalBook}. We reserve the superscript ${\bf{A}}^{T}$ for the transpose of ${\bf{A}}$, ${\bf{A}}^{\dagger}$ for the conjugate transpose of ${\bf{A}}$ and ${\bf{A}}^{-1}$ for the the pseudo-inverse ${\bf{A}}^{-1} = ({\bf{A}}^\dagger {\bf{A}})^{-1} {\bf{A}}^\dagger$, which reduces to the usual inverse if the number of columns and rows are same. The matrices ${\bf{I}}_i$ and ${\bf{1}}_{i \times j}$ denote the $i \times i$ identity and the $i \times j$ all-one matrix, respectively. The covariance matrix ${\bf{R}}_X$ of the random vector $X$ is computed ${\bf{R}}_X = E [ X X^\dagger ]$, the cross covariance matrix ${\bf{R}}_{XY}$ of $X$ and $Y$ is ${\bf{R}}_{XY} = E[ X Y^\dagger ]$, and ${\bf{R}}_{X|Y}$ denotes the conditional covariance matrix of $X$ conditioned on $Y$, i.e., ${\bf{R}}_{X|Y} = {\bf{R}}_{X} - {\bf{R}}_{XY} {\bf{R}}_Y^{-1} {\bf{R}}_{XY}^\dagger$. The covariance matrix ${\bf{R}}_Z$ of a matrix ${\bf{Z}}$ is denoted by ${\bf{R}}_Z = E [\textrm{vec}({\bf{Z}})\textrm{vec}({\bf{Z}})^\dagger]$. For a subset $\mathcal{S} \subseteq \{1, \dots, n\}$, given matrices ${\bf{X}}_1, \dots, {\bf{X}}_n$, we define the matrix $ {\bf{X}}_{\mathcal{S}}$ by stacking the matrices ${\bf{X}}_i$ with $i \in \mathcal{S}$ vertically in ascending order, namely $ {\bf{X}}_{\mathcal{S}} = \left [{\bf{X}}_1^T, \dots, {\bf{X}}_n^T \right]^T$.
\section{System Model}
\label{Sec:SM}
\begin{figure}[t]
\centering
\includegraphics[width=15cm]{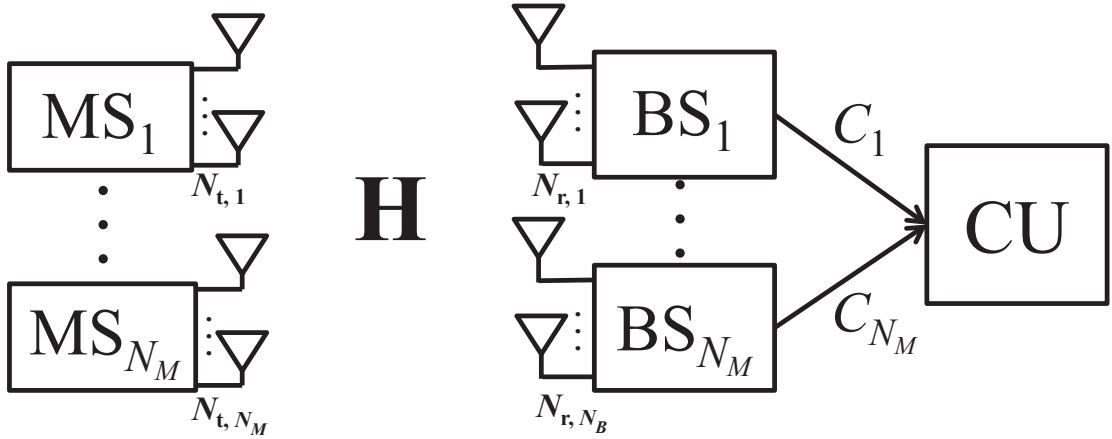}
\caption{System model.}
\label{fig:fig1}
\end{figure}
Consider the uplink of a cellular system consisting of $N_M$ MSs, $N_B$ BSs and a CU, as shown in Fig. \ref{fig:fig1}. We denote the set of all MSs as $\mathcal{N}_M = \{1, \dots, N_M\}$ and of all BSs as $\mathcal{N}_B = \{1, \dots, N_B\}$. The MSs, the $i$-th of which has $N_{t,i}$ transmit antennas, communicate in the uplink to the BSs, where the $j$-th BS is equipped with $N_{r,j}$ receive antennas. Each $j$-th  BS is connected to the CU via a backhaul link of capacity $C_j$. All rates, including $C_j$, are normalized to the bandwidth available on the uplink channel from MSs to BSs and are measured in bits/s/Hz. More precisely, we assume that $C_j T B$ bits can be transmitted on the backhaul by any $j$-th BS over an arbitrary number $B$ of coherence blocks. Note that each $j$-th BS can thus allocate its backhaul bits across different coherence blocks. This is akin to the standard long-term power constraints considered in a large part of the literature on fading channels (see, e.g., \cite{Caire99TIT}). We define $N_{\min} = \min (N_t, N_r)$ and $N_{\max}  = \max (N_t, N_r)$ where $N_t$ and $N_r$ are the number of total transmit antennas and total receive antennas, that is $N_t = \sum_{i=1}^{N_M} N_{t,i}$ and $N_r = \sum_{j=1}^{N_B} N_{r,j}$, respectively.

The channel coherence block, of length $T$ channel uses, is split it into a phase for channel training of length $T_p$ channel uses and a phase for data transmission of length $T_d$ channel uses, with 
\begin{equation}
T_p + T_d = T, \label{TC}
\end{equation}
as in \cite{Kobayashi11TSP, Hassibi03TIT, Zheng02TIT, Kobayashi11TCOM}. The signal transmitted by the $i$-th MS is given by a $N_{t,i} \times T$ complex matrix ${\bf{X}}_i$, where each column corresponds to the signal transmitted by the $N_{t,i}$ antennas in a channel use. This signal is divided into the $N_{t,i} \times T_p$ pilot signal ${\bf{X}}_{p,i}$ and the $N_{t,i} \times T_d$ data signal ${\bf{X}}_{d,i}$. We assume that the transmit signal ${\bf{X}}_i$ has a total per-block power constraint $\frac{1}{T}\left\| {\bf{X}}_i \right\|^2 = P_i$, and we define $\frac{1}{T_p}\left\| {\bf{X}}_{p,i} \right\|^2 = P_{p,i}$ and $\frac{1}{T_d}\left\| {\bf{X}}_{d,i} \right\|^2  = P_{d,i}$ as the powers used for training and data, respectively by the $i$-th MS. In terms of pilot and data signal powers, then, the power constraint becomes 
\begin{equation}
 \frac{T_p }{T}P_{p,i}  + \frac{T_d }{T} P_{d,i}  = P_i. \label{PC}
\end{equation} 
For simplicity, we assume equal transmit power allocation for each antenna of all MSs, and hence we have $P_i = P$, $P_{d,i} = P_d$ and $P_{p,i}=P_p$ for all $i \in \mathcal{N}_M$. We define ${\bf{X}}_p$ and ${\bf{X}}_d$ as the overall pilot signal and the data signal transmitted by all MSs, respectively, i.e., ${\bf{X}}_p = [{\bf{X}}_{p,1}^T, \dots, {\bf{X}}_{p,N_M}^T]^T$ and ${\bf{X}}_d = [{\bf{X}}_{d,1}^T, \dots, {\bf{X}}_{d,N_M}^T]^T$.

As in \cite{Kobayashi11TSP, Hassibi03TIT}, we assume that coding is performed across multiple channel coherence blocks. This implies that the ergodic capacity describes the system performance in terms of achievable sum-rate. Moreover, the training signal is ${\bf{X}}_p = {\sqrt{\frac{P_p}{N_t}}} {\bf{S}}_p$ where ${\bf{S}}_p$ is a $N_t \times T_p$ matrix of i.i.d. $\mathcal{CN}(0,1)$ variables. This implies that an independently generated training sequence with power $P_p/N_t$ is transmitted from each transmitting antenna across all MSs. Similarly, during the data phase, the MSs transmit independent streams with power $P_d/N_t$ from its transmitting antennas using spatial multiplexing. As a result, we have ${\bf{X}}_d = {\sqrt{\frac{P_d}{N_t}}} {\bf{S}}_d$ where ${\bf{S}}_d$ is a $N_t \times T_d$ matrix of i.i.d. $\mathcal{CN}(0,1)$ variables.

The $N_{r,j} \times T$ signal ${\bf{Y}}_j$ received by the $j$-th BS in a given coherence block, where each column corresponds to the signal received by the $N_{r,j}$ antennas in a channel use, can be split into the $N_{r,j} \times T_p$ received pilot signal ${\bf{Y}}_{p,j}$ and the $N_{r,j} \times T_d$ data signal ${\bf{Y}}_{d,j}$. The received signal at the $j$-th BS is then given by
\begin{subequations}
\label{RS;BS}
\begin{eqnarray}
\label{RS_p;BS} {\bf{Y}}_{p,j} &=& \sqrt{\frac{P_p}{N_t}} {\bf{H}}_j {\bf{S}}_p  + {\bf{Z}}_{p,j} \\
\label{RS_d;BS} \textrm{and}\,\,\,\, {\bf{Y}}_{d,j} &=& \sqrt{\frac{P_d}{N_t}} {\bf{H}}_j {\bf{S}}_d  + {\bf{Z}}_{d,j},
\end{eqnarray}
\end{subequations}
where ${\bf{Z}}_{p,j}$ and ${\bf{Z}}_{d,j}$ are respectively the $N_{r,j} \times T_p$ and $N_{r,j} \times T_d$ matrices of independent and identically distributed (i.i.d.) complex Gaussian noise variables with zero-mean and unit variance, i.e, $\mathcal{CN}(0,1)$. The $N_{r,j} \times N_t$ channel matrix ${\bf{H}}_j$ collects all the $N_{r,j} \times N_{t,i}$ channel matrix ${\bf{H}}_{ji}$ from the $i$-th MS to the $j$-th BS as ${\bf{H}}_j = [{\bf{H}}_{j1}, \dots, {\bf{H}}_{jN_M}]$. 

The channel matrix ${\bf{H}}_{ji}$ is modeled as Rician fading with the line-of-sight (LOS) component $\bar {\bf{H}}_{ji}$, which is deterministic, and the scattered component ${\bf{H}}_{w,ji}$ with i.i.d. $\mathcal{CN}(0,1)$ entries. Overall, the channel matrix ${\bf{H}}_{ji}$ between the $j$-th BS and the $i$-th MS is represented as
\begin{eqnarray} \label{Rician channel}
{\bf{H}}_{ji} = \sqrt{\alpha}_{ji} \left( \sqrt{\frac{K}{K+1}} \bar {\bf{H}}_{ji} + \sqrt{\frac{1}{K+1}} {\bf{H}}_{w,ji}\right),
\end{eqnarray}
where the Rician factor $K$ defines the power ratio of the LOS component and the scattered component, and the parameter $\alpha_{ji}$ represents the power gain between the $j$-th BS and the $i$-th MS. The channel matrix ${\bf{H}}_j$ is assumed to be constant during each channel coherence block and to change according to an ergodic process from  block to block.

\section{Preliminaries}
\label{Sec:Pre} In this section, we discuss two reference schemes. The first is a non-coherent strategy, whereby the MSs do not transmit any pilot signal (i.e., $T_p=0$), each $j$-th BS compresses its received data signal (\ref{RS_d;BS}) for transmission on the backhaul, and the CU performs non-coherent decoding \cite{Marzetta99TIT}. The second approach is the CFE strategy first studied in \cite{Kobayashi11TSP}, whereby each $j$-th BS compresses and transmits also its received pilot signals (\ref{RS_p;BS}); the CU estimates the CSI based on the compressed pilot signals received on the backhaul links; and the estimated CSI is then used by the CU to perform coherent decoding. To simplify the presentation, in this section, we assume a single BS, i.e., $N_B=1$, and hence drop the BS index $j$. Additionally, in non-coherent processing, we assume a single MS and drop the MS index $i$.

\subsection{Non-Coherent Processing} 
\label{subSec:NCP_Pre} With non-coherent processing, the MS transmits the data signals ${\bf{X}}_d$ during the entire channel coherence time $T$ (i.e., $T_p = 0$). The BS compresses the vector of received signals ${\bf{Y}}_d$ (\ref{RS_d;BS}) across all coherence times in the coding block and sends it to the CU on the backhaul link. Accordingly, the compressed received signals $\widehat {\bf{Y}}_d$ available at the CU can be written as 
\begin{equation}
\widehat {\bf{Y}}_{d} = {\bf{Y}}_{d}  + {\bf{Q}}_{d}, \label{RDS;BS_NCP}
\end{equation}
where ${\bf{Q}}_{d}$ is independent of ${\bf{Y}}_{d}$ and represents the quantization noise matrix, which is assumed for simplicity to have i.i.d. 
$\mathcal{CN}(0,\sigma_d^2)$ entries. 

\rem \label{IIDQuanNoise} It is noted that, in principle, the design of the quantizers could be adapted to the channel statistics. Here, and in most of the paper, we instead assume i.i.d. quantization noises. Beside simplifying the system design, this choice is known to be optimal in the high-resolution regime (see the discussion on reverse waterfilling in \cite[Ch. 10]{CoverBook}). Another advantage of independent compression noises is that, if the signals to be compressed are not too correlated, then close-to-optimal quantization can be obtained with a separate quantizer for each component\footnote{Independent signals can be in fact optimally compressed by separate quantizers, as it can be seen from the fact that the rate-distortion function for a set of independent signals can be written as the sum of the individual rate-distortion functions (see \cite[Ch. 10]{CoverBook}).}. $\hspace{0.3cm} \blacksquare$

Using standard rate-distortion theoretic arguments, the quantization noise $\sigma_d^2$ depends on the backhaul capacity via the equation $I({\bf{Y}}_d; \widehat {\bf{Y}}_d) = C$, which leads to $\sigma_d^2 = (1+P)/(2^{(C/N_r)} -1)$ (see, e.g., \cite[Ch. 3]{GamalBook}). A lower bound on the capacity achievable with non-coherent decoding can be obtained by substituting the equivalent SNR $\rho=P/(1+\sigma^2_d)$ in \cite[Eq. (10)]{Marzetta99TIT}\footnote{It is remarked that this rate is achieved by choosing the codewords ${\bf{X}}_d$ according to an appropriate orthogonal signaling scheme \cite{Marzetta99TIT} and not via Gaussian random codebooks as described in Section \ref{Sec:SM} and assumed in the rest of the paper.}.

\subsection{Compress-Forward-Estimate (CFE)}
\label{subSec:CFE_Pre} With the CFE scheme, the BS compresses both its received pilot signal (\ref{RS_p;BS}) and its received data signal (\ref{RS_d;BS}), and forwards them to the CU on the backhaul link. The CU then estimates the CSI based on the received compressed pilot signals and performs coherent decoding. 

\subsubsection{Training Phase}
During the training phase, the vector of received training signals $ {\bf{Y}}_{p}$ (\ref{RS_p;BS}) across all coherence times is compressed as
\begin{equation}
\widehat {\bf{Y}}_{p} = {\bf{Y}}_{p}  + {\bf{Q}}_{p}, \label{CTS;CU}
\end{equation}
where the compression noise matrix ${\bf{Q}}_{p}$ is assumed to have i.i.d.  $\mathcal{CN} (0, \sigma_{p}^2)$ entries (see Remark \ref{IIDQuanNoise}). Based on (\ref{CTS;CU}), the channel matrix ${\bf{H}}_i$ from $i$-th MS to the BS is estimated at the CU by the minimum mean square error (MMSE) method. Hence, it can be expressed as 
\begin{equation}
{\bf{H}}_i = \widehat {\bf{H}}_i + {\bf{E}}_i, \label{ECS+EE;CU}
\end{equation}
where the estimated channel $\widehat {\bf{H}}_i$ is a complex Gaussian matrix with mean matrix $\sqrt{\frac{\alpha_{i} K}{K+1}} \bar{\bf{H}}_i$ and covariance matrix $\sigma_{\widehat h_i}^2 {\bf{I}}_{N_r N_{t,i}}$, and the estimation error ${\bf{E}}_i$ has i.i.d. $\mathcal{CN} (0, \sigma_{e_i}^2)$ entries. The variances of the estimated channel and the estimation error can be calculated as $\sigma_{\widehat h_i}^2 = \frac{{\frac{\alpha_i}{K+1}} T_p P_p}{{T_p P_p +  N_t (1+\sigma_{p}^2) ({K+1})}}$ and $\sigma_{e_i}^2 = \frac{ \alpha_i N_t (1+\sigma_{p}^2)}{{T_p P_p +  N_t (1+\sigma_{p}^2) ({K+1})}}$, respectively (see, e.g., \cite{Hassibi03TIT, Bjornson10TSP}).

\subsubsection{Data Phase}
The compressed data signal received at the CU in (\ref{RDS;BS_NCP}) can be written as the sum of a useful term $\widehat {\bf{H}} {\bf{X}}_d$ and of the equivalent noise ${\bf{N}}_{d} = {\bf{E}} {\bf{X}}_d + {\bf{Z}}_{d} + {\bf{Q}}_{d}$, namely 
\begin{equation}
\widehat {\bf{Y}}_{d} = \widehat {\bf{H}} {\bf{X}}_d + {\bf{N}}_{d}, \label{RDS;CU:ComForEst}
\end{equation}
where the equivalent noise ${\bf{N}}_{d}$ has zero-mean and covariance matrix 
\begin{eqnarray}
{\bf{R}}_{N} = E [{\textrm{vec}}( {\bf{N}}_{d}) {\textrm{vec}}({\bf{N}}_{d})^\dagger] 
= \left ( 1 + \sigma_{d}^2 + \frac{P_d}{N_t} \sum_{i=1}^{N_M} N_{t,i} \sigma_{e_{i}}^2 \right ) {\bf{I}}_{N_{r} T_d}. \label{CM_TN:ComForEst}
\end{eqnarray}

\subsubsection{Ergodic Achievable Rate}
The ergodic capacity is given by the mutual information $\frac{1}{T} I ( {\bf{X}}_d; \widehat {\bf{Y}}_d | \widehat {\bf{H}} )$ [bits/s/Hz] (see, e.g, \cite[Ch. 3]{GamalBook}), which is bounded in the next lemma. 
\lemma \label{lemma;Pre_CFE} Let $C_{p}$ and $C_{d}$ define the backhaul rates allocated respectively to the compressed pilot and data signals on the backhaul from the BS to the CU. The ergodic capacity for the CFE strategy can be bounded as $\frac{1}{T} I ( {\bf{X}}_d; \widehat {\bf{Y}}_d | \widehat {\bf{H}} ) \ge R$, where
\begin{eqnarray}
R = \frac{T_d}{T} E \left [ \log_2 \det \left( {\bf{I}}_{N_r} + \rho_{\textrm{eff}} \widehat {\bf{H}} \widehat {\bf{H}}^\dagger \right) \right], \label{EMI;ComForEst}
\end{eqnarray}
with $\rho_{\textrm{eff}} = \frac{P_d}{N_t \left(1 + \sigma_d^2 + \frac{P_d}{N_t} \sum_{i=1}^{N_M} N_{t,i} \sigma_{e_{i}}^2 \right)}$, and $\widehat {\bf{H}}$ being distributed as in (\ref{ECS+EE;CU}). Moreover, the quantization noise powers $(\sigma_p^2, \sigma_d^2)$ must satisfy the backhaul constraint $C_p + C_d = C$, where 
\begin{subequations}
\label{BR;CFE}
\begin{eqnarray}
\label{BR;CFE_DS} C_d &=& \frac{T_d}{T} \log_2 \det \left( {\bf{I}}_{N_r} + \frac{\frac{P_d}{N_t} \left( \frac{K}{K+1} \bar {\bf{H}} \bar{\bf{H}}^\dagger + \frac{\sum_{i=1}^{N_M} \alpha_{i} N_{t,i}}{K+1} {\bf{I}}_{N_r} \right) +{\bf{I}}_{N_r} }{\sigma_{d}^2} \right) \\
\label{BR;CFE_TS}  \textrm{and} \hspace{0.5cm}  C_p &=& \frac{T_p}{T} \log_2 \det \left( {\bf{I}}_{N_r} + \frac{\frac{P_p}{N_t} \left( \frac{K}{K+1} \bar {\bf{H}} \bar{\bf{H}}^\dagger + \frac{\sum_{i=1}^{N_M} \alpha_{i} N_{t,i}}{K+1} {\bf{I}}_{N_r} \right) +{\bf{I}}_{N_r} }{\sigma_{p}^2} \right).
\end{eqnarray}
\end{subequations}
\begin{IEEEproof}
Since a closed-form expression is not known, here we consider a lower bound obtained by overestimating the detrimental effect of the estimation error \cite{Hassibi03TIT, Medard00TIT, Weingarten04TIT}. This is done by treating the total noise term ${\bf{N}}_{d}$ in (\ref{RDS;CU:ComForEst}) as being independent of ${\bf{X}}_d$ and zero-mean complex Gaussian \cite{Hassibi03TIT, Medard00TIT, Weingarten04TIT}. The resulting lower bound $R$ can then be evaluated as (\ref{EMI;ComForEst}). Then, from standard rate-distortion theoretic considerations \cite[Ch. 3]{GamalBook}, we can relate the backhaul rates $C_d$ for data transmission with the variances of the compression noise $\sigma_d^2$ as 
\begin{eqnarray}
\nonumber C_{d}  &=& \frac{1}{T} I( {\bf{Y}}_{d}; \widehat {\bf{Y}}_{d}) \\
\nonumber &=& \frac{1}{T} \left( h({\bf{Y}}_{d} + {\bf{Q}}_{d} ) - h({\bf{Q}}_{d}) \right) \\
&\le& \frac{T_d}{T} \log_2 \det \left( {\bf{I}}_{N_r} + \frac{\frac{P_d}{N_t} \left( \frac{K}{K+1} \bar {\bf{H}} \bar{\bf{H}}^\dagger + \frac{\sum_{i=1}^{N_M} \alpha_{i} N_{t,i}}{K+1} {\bf{I}}_{N_r} \right) +{\bf{I}}_{N_r} }{\sigma_{d}^2} \right), \label{DBC;SC_IC}
\end{eqnarray}
where we have used the test channel defined by (\ref{RDS;BS_NCP}) and the maximum entropy theorem to bound the differential entropy $h ( {\bf{Y}}_{d} + {\bf{Q}}_{d})$ in the last line \cite{CoverBook}. Note that the upper bound (\ref{DBC;SC_IC}) overestimates the backhaul rate $C_d$ needed to convey the received data signal over the backhaul link. Therefore, its application leads to feasible solutions for the original problem. In (\ref{BR;CFE}), we make the conservative choice of imposing equality in (\ref{DBC;SC_IC}). In a similar manner, we obtain the relation between the quantization error variance $\sigma_p^2$ and the backhaul rate $C_p$ for training transmission as (\ref{BR;CFE_TS}).
\end{IEEEproof}

For the CFE scheme, the ergodic achievable sum-rate (\ref{EMI;ComForEst}) can now be optimized over the backhaul allocation $( C_{p}, C_{d} )$ under the backhaul constraint $C=C_p+C_d$, with $C_p$ and $C_d$ in (\ref{BR;CFE}), by maximizing the effective SNR $\rho_{\textrm{eff}}$ in (\ref{EMI;ComForEst}). This non-convex problem can be tackled using a line search method \cite{BoydBook} in a bounded interval (e.g., over $C_p$ in the interval $[0, C]$).

\rem \label{Rem:EAR} The lower bound $R$ on the ergodic capacity in (\ref{EMI;ComForEst}), and related bounds in the next section, will be referred thereafter as the \textit{ergodic achievable rate}. $\hspace{9.8cm} \blacksquare$
\section{Estimate-Compress-Forward (ECF)}
Here, we introduce the ECF approach. Accordingly, each BS estimates the CSI based on its received pilot signal (\ref{RS_p;BS}), and then compresses both its estimated CSI and its received data signal (\ref{RS_d;BS}) for transmission on the backhaul. In this section, we introduce the key common quantities that define the class of ECF schemes, which are then studied in Section \ref{Sec:SBC} for the single BS case and in Section \ref{Sec:MBC} for the more general multiple BSs case.

\subsection{Training Phase}
The MMSE estimate of ${\bf{H}}_j$ performed at the $j$-th BS given the observation ${\bf{Y}}_{p,j}$ in (\ref{RS_p;BS}) is given by
\begin{equation}
\widetilde {\bf{H}}_j = \sqrt{\frac{N_t}{P_p}} \bar{\bf{Y}}_{p,j} {\bf{S}}_p^\dagger \left( \frac{N_t \left(K+1\right)}{P_p} {\bf{I}}_{N_r} + {\bf{S}}_p {\bf{S}}_p^\dagger\right)^{-1} + \sqrt{\frac{K}{K+1}} \bar{\bf{H}}_j, \label{MMSE}
\end{equation}
where $\bar{\bf{Y}}_{p,j} = {\bf{Y}}_{p,j} - \sqrt{\frac{P_p}{N_t} \frac{K}{K+1}} \bar{\bf{H}}_j {\bf{S}}_p$ and $\bar {\bf{H}}_j = [\sqrt{\alpha}_{j1} \bar {\bf{H}}_{j1}, \dots, \sqrt{\alpha}_{jN_M} \bar {\bf{H}}_{jN_M}]$ (see, e.g., \cite{Hassibi03TIT, Bjornson10TSP}). The estimated channel $\widetilde {\bf{H}}_j = [\widetilde {\bf{H}}_{j1}, \dots, \widetilde {\bf{H}}_{jN_M}]$ in (\ref{MMSE}) is such that the estimated channel matrix $\widetilde {\bf{H}}_{ji}$ corresponding to the channel between the $j$-th BS and $i$-th MS has a matrix-variate complex Gaussian distribution with mean matrix $\sqrt{\frac{\alpha_{ji} K}{K+1}} \bar{\bf{H}}_{ji}$ and covariance matrix $\sigma_{\widetilde h_{ji}}^2 {\bf{I}}_{N_{r,j}}$, where $\sigma_{\widetilde h_{ji}}^2 = \frac{\frac{\alpha_{ji}}{K+1} T_p P_p}{{T_p P_p +  N_t (K+1)}}$. Moreover, we can decompose the channel matrix ${\bf{H}}_{ji}$ into the estimate $\widetilde {\bf{H}}_{ji}$ and the independent estimation error ${\bf{E}}_{ji}$, as 
\begin{equation}
{\bf{H}}_{ji} = \widetilde {\bf{H}}_{ji} + {\bf{E}}_{ji}, \label{ECS+EE}
\end{equation}
where the error ${\bf{E}}_{ji}$ has i.i.d. $\mathcal{CN} (0, \sigma_{e_{ji}}^2)$ entries with $\sigma_{e_{ji}}^2 =  \frac{\alpha_{ji} N_t}{{T_p P_p +  N_t (K+1)}}$.

The sequence of channel estimates $\widetilde {\bf{H}}_j$ for all coherence times in the coding block is compressed by the $j$-th BS and forwarded to the CU on the backhaul link. The compressed channel $\widehat {\bf{H}}_j$ is related to the estimate $\widetilde {\bf{H}}_j$ as
\begin{equation}
\widetilde {\bf{H}}_j = \widehat {\bf{H}}_j + {\bf{Q}}_{p,j}, \label{CCS+QN}
\end{equation}
where the $N_{r,j} \times N_t$ quantization noise matrix ${\bf{Q}}_{p,j}$ has zero-mean i.i.d. $\mathcal{CN}(0,\sigma_{p,j}^2)$ entries (see Remark \ref{IIDQuanNoise}) and the compressed estimate $\widehat {\bf{H}}_j$ is complex Gaussian with mean matrix $\sqrt{\frac{K}{K+1}} \bar{\bf{H}}_j$ and covariance matrix ${\bf{R}}_{\widetilde h_j}-\sigma_{p,j}^2 {\bf{I}}_{N_t}$, where ${\bf{R}}_{\widetilde h_j}$ is diagonal matrix with main diagonals given by $[\sigma_{\widetilde h_{j1}}^2 {\bf{I}}_{N_{t,1}}, \dots, \sigma_{\widetilde h_{jN_M}}^2 {\bf{I}}_{N_{t,N_M}}]$ (see, e.g., \cite[Ch. 3]{GamalBook}). We will discuss in Section \ref{Sec:SBC} and Section \ref{Sec:MBC} how to relate the quantization noise variance $\sigma_{p,j}^2$ to the backhaul capacity $C_j$.

\subsection{Data Phase}
\label{Sec:DP;SM} During the data phase, the $j$-th BS compresses the signal ${\bf{Y}}_{d,j}$ in (\ref{RS_d;BS}) and sends it to the CU on the backhaul link. The received signals at the CU are related to ${\bf{Y}}_{d,j}$ as
\begin{equation}
\widehat {\bf{Y}}_{d,j} = {\bf{Y}}_{d,j}  + {\bf{Q}}_{d,j}, \label{RDS;BS}
\end{equation}
where ${\bf{Q}}_{d,j}$ is independent of ${\bf{Y}}_{d,j}$ and represents the quantization noise matrix\footnote{Note that we use a different formulation for the quantization test channel (see, e.g., \cite[Ch. 3]{GamalBook}) in (\ref{RDS;BS}) with respect to (\ref{CCS+QN}). In (\ref{RDS;BS}) and similarly in (\ref{RDS;BS_NCP}) and (\ref{CTS;CU}), in fact, the quantization noise is added to the signal to be compressed. While the formulation in (\ref{CCS+QN}) is optimal from a rate-distortion point of view \cite[Ch. 3]{GamalBook}, the test channel (\ref{RDS;BS}) is selected here for its analytical convenience. It is noted that this test channel is assumed in many previous studies, including \cite{Sanderovich09TIT, Zhou12arXiv, Kobayashi11TSP, Lim11TIT}.}. This is assumed to be zero-mean complex Gaussian with covariance matrix $E [{\textrm{vec}}( {\bf{Q}}_{d,j}) {\textrm{vec}}({\bf{Q}}_{d,j})^\dagger] = {\bf{R}}_{d,j} \otimes {\bf{I}}_{T_d}$. By this definition, ${\bf{R}}_{d,j}$ is the covariance matrix of the $N_{r,j} \times 1$ compression noise vector for all the channel uses in a data transmission period. Following our design choices for the other quantization noises, we will mostly assume ${\bf{R}}_{d,j}$ to be a scaled identity matrix, namely ${\bf{R}}_{d,j} = \sigma_{d,j}^2 {\bf{I}}_{N_{r,j} T_d}$ (see Remark \ref{IIDQuanNoise}). However, we will allow this covariance matrix to be arbitrary in Section \ref{Sec:JAC;SBC} in order to illustrate the potential advantages of a system design that adapts the quantizers to the current channel conditions (see also Remark \ref{IIDQuanNoise}). The relationship of matrix ${\bf{R}}_{d,j}$ with the backhaul capacity will be clarified in the next sections. 

We close this section by deriving a model for the received signals at the CU that is akin to (\ref{RDS;CU:ComForEst})-(\ref{CM_TN:ComForEst}) for CFE. With ECF, the CU recovers the sequence of quantized data signals $\widehat {\bf{Y}}_{d,j}$ in (\ref{RDS;BS}) and of quantized channel estimates $\widehat {\bf{H}}_j$ in (\ref{CCS+QN}) from the information received on the backhaul link. Separating the desired signal and the noise in (\ref{RDS;BS}), the received signal $\widehat {\bf{Y}}_{d,j}$ from the $j$-th BS can be expressed as
\begin{eqnarray}
\widehat {\bf{Y}}_{d,j} = \widehat {\bf{H}}_j {\bf{X}}_d + {\bf{N}}_{d,j}, \label{RDS;CU}
\end{eqnarray}
where ${\bf{N}}_{d,j}$ denotes the equivalent noise ${\bf{N}}_{d,j} = \left( {\bf{Q}}_{p,j} + {\bf{E}}_j \right) {\bf{X}}_d + {\bf{Z}}_{d,j} + {\bf{Q}}_{d,j}$, which has zero-mean and covariance matrix
\begin{eqnarray}
{\bf{R}}_{N_j} = E [{\textrm{vec}}( {\bf{N}}_{d,j}) {\textrm{vec}}({\bf{N}}_{d,j})^\dagger] =  {\bf{R}}_{d,j} \otimes {\bf{I}}_{T_d} + \sigma_{{pe}_j}^2 {\bf{I}}_{N_{r,j} T_d} \label{CM_TN}
\end{eqnarray}
with 
\begin{equation}
\label{Noise_pe} \sigma_{pe, j}^2 = \left ( 1 + P_d \left( \sigma_{p,j}^2 + \frac{\sum_{i=1}^{N_M} N_{t,i} \sigma_{e_{ji}}^2}{N_t} \right) \right ),
\end{equation}
where we have used the relations $E[{\bf{Q}}_{p,j} {\bf{Q}}_{p,j}^\dagger] = N_t \sigma_{p,j}^2 {\bf{I}}_{N_{r,j}}$ and $E[{\bf{E}}_j {\bf{E}}_j^\dagger] = \sum_{i=1}^{N_M} N_{t,i} \sigma_{e_{ji}}^2 {\bf{I}}_{N_{r,j}}$. We observe that, as in (\ref{RDS;CU:ComForEst})-(\ref{CM_TN:ComForEst}), ${\bf{N}}_{d,j}$ is not Gaussian distributed and is not independent of ${\bf{X}}_d$ (see also \cite{Hassibi03TIT}).
\section{Analysis of ECF : The Single Base Station Case}
\label{Sec:SBC} In this section, we discuss how to calculate the compression noises statistics, namely $\sigma_{p,j}^2$ for the estimated CSI (see (\ref{CCS+QN})) and ${\bf{R}}_{d,j}$ for the data (see (\ref{RDS;BS})). We consider three different strategies in order of complexity, namely separate compression, joint compression and joint adaptive compression of estimated CSI and received data signal. Specifically, here, we first consider the single base station case, i.e., $N_B=1$. The more complex scenario with multiple BSs will be studied in Section \ref{Sec:MBC} by building on the analysis in this section. For simplicity of notation, we drop the BS index in this section.


\subsection{Separate Compression of Channel and Received Data Signal}
\label{Sec:SC;SBC} Here, we consider the conventional option of compressing separately the sequence of the estimated channels $\widetilde {\bf{H}}$ and of the received data signals ${\bf{Y}}_d$. For simplicity, and due to the identical distribution of the entries of ${\bf{Y}}_{d}$, here we choose ${\bf{R}}_{d} = \sigma_{d}^2 {\bf{I}}_{N_{r}}$ (see Remark 1). 
\prop Let $C_p$ and $C_d$ denote respectively the backhaul rates allocated for the transmission of the compressed channel estimates (\ref{CCS+QN}) and of the compressed received signals (\ref{RDS;BS}) on the backhaul link from the BS to the CU. The ergodic achievable sum-rate for separate compression strategy is given as
\begin{eqnarray}
R = \frac{T_d}{T} E \left[ \log _2 \det \left( {\bf{I}}_{N_r} + \rho_{{\textrm{eff}}} { \widehat {\bf{H}}} {\widehat {\bf{H}}}^\dagger \right) \right], \label{EAR;SC_IC}
\end{eqnarray}
with 
\begin{equation}
\rho_{{\textrm{eff}}}  = \frac{P_d}{ N_t \left( 1 + \sigma_{d}^2 + P_d \left( \sigma_{p}^2 + {\sum_{i=1}^{N_M} N_{t,i} \sigma_{e_{i}}^2}/{N_t} \right) \right)} \label{ESNR;SC_IC},
\end{equation} 
with $\widehat {\bf{H}}$ being distributed as in (\ref{CCS+QN}), and with $\sigma_{e_i}^2$ in (\ref{ECS+EE}). Moreover, the quantization noise powers $(\sigma_p^2, \sigma_d^2)$ must satisfy the backhaul constraint $C_p + C_d = C$, where 
\begin{subequations}
\label{BR;ECF_Sep}
\begin{eqnarray}
\label{BR;ECF_Sep_DS} C_p &=& \frac{N_r}{T} \log_2 \left( \frac{ \prod_{i=1}^{N_M} \left( \sigma_{\widetilde h_{i}}^2\right)^{N_{t,i}}}{(\sigma_{p}^2)^{N_t}} \right) \\
\label{BR;ECF_Sep_TS}  \textrm{and} \hspace{0.5cm}  C_d &=& \frac{T_d}{T} \log_2 \det \left( {\bf{I}}_{N_r} + \frac{\frac{P_d}{N_t} \left( \frac{K}{K+1} \bar {\bf{H}} \bar{\bf{H}}^\dagger + \frac{\sum_{i=1}^{N_M} \alpha_{i} N_{t,i}}{K+1} {\bf{I}}_{N_r} \right) +{\bf{I}}_{N_r} }{\sigma_{d}^2} \right),
\end{eqnarray}
\end{subequations}
with $\sigma_{\widetilde h_{i}}^2$ being given in (\ref{MMSE}). 
\begin{IEEEproof}
As in the proof of Lemma \ref{lemma;Pre_CFE}, a lower bound on the ergodic achievable sum-rate is obtained by overestimating the detrimental effect of the estimation error, and the resulting ergodic achievable sum-rate $R$ can be evaluated as in (\ref{EAR;SC_IC}). Then, from standard rate-distortion theoretic considerations \cite{CoverBook}, we can relate the compression noise power $\sigma_{p}^2$ with the backhaul capacity needed for the transmission of the sequence of channel estimates $\widehat {\bf{H}}$ as
\begin{equation}
\label{CBC;SC_IC} C_{p}  = \frac{1}{T} I( \widetilde {\bf{H}}; \widehat {\bf{H}}) = \frac{1}{T} \left( h(\widehat {\bf{H}} + {\bf{Q}}_{p} ) - h({\bf{Q}}_{p}) \right) = \frac{N_r}{T} \log_2 \left( \frac{ \prod_{i=1}^{N_M} \left( \sigma_{\widetilde h_{i}}^2\right)^{N_{t,i}}}{(\sigma_{p}^2)^{N_t}} \right),
\end{equation}
where we have used the test channel defined by (\ref{CCS+QN}). It follows that the CSI quantization noise is
\begin{equation}
\sigma _{p}^2=\left(\prod_{i=1}^{N_M} \left( \sigma_{\widetilde h_{i}}^2\right)^{N_{t,i}}\right)^{\frac{1}{N_t}} 2^{ - T C_{p} / (N_{r} N_t) }. \label{CV_CQN}
\end{equation} 
Moreover, equation (\ref{BR;ECF_Sep_TS}) follows in the same way as (\ref{DBC;SC_IC}).
\end{IEEEproof}

As for CFE, the ergodic achievable sum-rate (\ref{EAR;SC_IC}) can now be optimized over the backhaul allocation $( C_{p}, C_{d} )$ under the backhaul constraint $C=C_p+C_d$, with $C_p$ and $C_d$ in (\ref{BR;ECF_Sep}), by maximizing the effective SNR $\rho_{\textrm{eff}}$ in (\ref{ESNR;SC_IC}) using a line search \cite{BoydBook} in a bounded interval.

\rem \label{rem:SBC} If we consider the special case of a Rayleigh fading channel, that is $K=0$, the ergodic achievable sum-rate (\ref{EAR;SC_IC}) can be evaluated explicitly following \cite{Shin03TIT}. Moreover, by imposing equality in (\ref{BR;ECF_Sep_TS}), we can easily calculate the quantization variance $\sigma_d^2$ as
\begin{equation}
\sigma_{d}^2 = \frac{\frac{P_d}{N_t} \sum_{i=1}^{N_M} \alpha_i N_{t,i}+1}{{2^{T C_{d} /{(N_{r} T_d )}}  - 1}}. \label{CV_DQN} \vspace{-0.6cm}
\end{equation} 
$\hspace{17.8cm} \blacksquare$

\rem For Rayleigh fading ($K=0$) and $N_r=N_t=1$, the ergodic achievable sum-rate (\ref{EMI;ComForEst}) obtained with CFE equals the ergodic achievable sum-rate (\ref{EAR;SC_IC}) with ECF based on separate compression. Further comparisons among the discussed methods will be presented in Section \ref{Sec:Numerical Results} via numerical results. $\hspace{4.5cm} \blacksquare$

\rem In the discussion above, we have considered the power allocation $(P_p, P_d)$ and the time allocation $(T_p, T_d)$ as fixed. The optimization of these parameters can be carried out similar to \cite{Hassibi03TIT} and is not further detailed here. $\hspace{17cm} \blacksquare$
\subsection{Joint Compression of Channel and Received Data Signal}
\label{Sec:JC;SBC} Here we propose a more sophisticated method to convey the sequence of the channel estimates $\widehat {\bf{H}}$ in (\ref{CCS+QN}) and of received data signals $\widehat {\bf{Y}}_{d}$ in (\ref{RDS;BS}) over the backhaul link. This method leverages the fact that channel estimates $\widetilde {\bf{H}}$ in (\ref{ECS+EE}) and received signals ${\bf{Y}}_{d}$ in (\ref{RS_d;BS}), and thus $\widehat {\bf{H}}$ and $\widehat {\bf{Y}}_{d}$, are correlated. As in Section \ref{Sec:SC;SBC}, we assume an uncorrelated compression covariance ${\bf{R}}_{d} = \sigma_{d}^2 {\bf{I}}_{N_{r}}$ in (\ref{RDS;BS}) and we are interested in finding the optimal pair $(\sigma_p^2, \sigma_d^2)$. 

\prop The ergodic achievable sum-rate for joint compression strategy can be bounded as (\ref{EAR;SC_IC}), where $\rho_{\textrm{eff}}$ is given by (\ref{ESNR;SC_IC}). Moreover, the quantization noise powers $(\sigma_p^2, \sigma_d^2)$ must satisfy the backhaul constraint $C_p + C_d = C$, where
\begin{eqnarray}
C_d = \frac{T_d}{T} \left( E \left [ \log_2 \det \left( {\bf{I}}_{N_r} + \rho_{{\textit{eff}}} \widehat {\bf{H}} \widehat {\bf{H}}^\dagger \right)\right] + N_{r} \log_2 \left(\sigma_{pe}^2 + \sigma_{d}^2 \right)  - N_{r} \log_2 \sigma_{d}^2 \right), \label{BC;JC_IC}
\end{eqnarray}
and $C_p$ is defined in (\ref{CBC;SC_IC}), with $\widehat {\bf{H}}$ being distributed as in (\ref{CCS+QN}) and $\sigma_{pe}^2$ being given in (\ref{Noise_pe}).
\begin{IEEEproof}
We only need to derive (\ref{BC;JC_IC}). To this end, from standard rate-distortion arguments, we have that the rate required on the backhaul is 
\begin{equation}
C = \frac{1}{T} I \left( {\bf{Y}}_{d},\widetilde {\bf{H}};\widehat {\bf{Y}}_{d},\widehat {\bf{H}} \right) = \frac{1}{T} \left( I \left( \widetilde {\bf{H}}; \widehat {\bf{H}} \right) + I \left( {\bf{Y}}_{d}; \widehat {\bf{Y}}_{d} | \widehat {\bf{H}} \right) \right), \label{BC_MI;JC_IC}
\end{equation}
where the second equality is shown in Appendix \ref{Appendix:ProofBC;JC}. As also shown in Appendix \ref{Appendix:ProofBC;JC}, equality (\ref{BC_MI;JC_IC}) implies the condition $C = C_p + C_d$, with $C_p$ in (\ref{CBC;SC_IC}) and $C_d$ in (\ref{BC;JC_IC}).
\end{IEEEproof}

The ergodic achievable sum-rate (\ref{EAR;SC_IC}) can now be optimized over the quantization noise powers $(\sigma_p^2, \sigma_d^2)$ under the backhaul constraint $C = C_p + C_d$, with $C_p$ in (\ref{CBC;SC_IC}) and $C_d$ in (\ref{BC;JC_IC}), using a two-dimensional search.

\rem It is useful to compare the backhaul constraint in (\ref{BR;ECF_Sep}), corresponding to separate compression, with $C = C_p + C_d$, which applies to joint compression with $C_p$ in (\ref{CBC;SC_IC}) and $C_d$ in (\ref{BC;JC_IC}). To this end, we observe that (\ref{BR;ECF_Sep}) can be expressed in terms of the quantization noise variance $\sigma_{p}^2$ and $\sigma_{d}^2$ using (\ref{CBC;SC_IC}) and (\ref{BR;ECF_Sep_TS}), leading to the condition
\begin{eqnarray}
C = C_p + \frac{T_d}{T} \log_2 \det \left( {\bf{I}}_{N_r} + \frac{\frac{P_d}{N_t} \left( \frac{K}{K+1} \bar {\bf{H}} \bar{\bf{H}}^\dagger + \frac{\sum_{i=1}^{N_M} \alpha_{i} N_{t,i}}{K+1} {\bf{I}}_{N_r} \right) +{\bf{I}}_{N_r} }{\sigma_{d}^2} \right). \label{CV_BC;SC_IC}
\end{eqnarray}
The difference between (\ref{CV_BC;SC_IC}) and the condition $C = C_p + C_d$, with $C_p$ in (\ref{CBC;SC_IC}) and $C_d$ in (\ref{BC;JC_IC}), is given as
\begin{eqnarray}
\frac{T_d}{T} \left(\log_2 \det \left( {\bf{I}}_{N_r} + \rho_{\textit{eff}} \left( \frac{K}{K+1} \bar {\bf{H}} \bar{\bf{H}}^\dagger + \left(\sum_{i=1}^{N_M}  N_{t,i} \sigma_{\widetilde h_i}^2 - N_t \sigma_p^2 \right) {\bf{I}}_{N_r} \right) \right) - E \left[ \log_2 \det \left( {\bf{I}}_{N_r} + \rho_{\textit{eff}} \widehat {\bf{H}} \widehat {\bf{H}}^\dagger \right) \right] \right) \ge 0, \label{D_SC&JC}
\end{eqnarray}
where the latter condition follows by Jensen's inequality since we have $E \left[ \widehat {\bf{H}} \widehat {\bf{H}}^\dagger\right] = \frac{K}{K+1} \bar {\bf{H}} \bar{\bf{H}}^\dagger + (\sum_{i=1}^{N_M}  N_{t,i} \sigma_{\widetilde h_i}^2 - N_t \sigma_p^2 ) {\bf{I}}_{N_{r}}$. Inequality (\ref{D_SC&JC}) shows that joint compression has the potential of improving the efficiency of backhaul utilization. This will be further explored via numerical results in Section \ref{Sec:Numerical Results}. $\hspace{5.4cm} \blacksquare$
\subsection{Joint Adaptive Compression of Channel and Received Data Signal}
\label{Sec:JAC;SBC} In this section, we introduce an improved method for joint compression of channel and received data signal. The main idea is that of adapting the covariance matrix ${\bf{R}}_{d}$ of the compression noise added to the data signal (see (\ref{RDS;BS})) to the channel estimate in each channel coherence block. The rationale for this approach is that if, e.g., the channel quality in a coherence block is poor, there is no reason to invest significantly backhaul capacity for the compression of the corresponding received data signal. We recall that, in the strategy studied in the previous section, the covariance matrix ${\bf{R}}_{d}$ was instead selected to be equal for all the coherence blocks (and given as ${\bf{R}}_d = \sigma_d^2 {\bf{I}}_{N_r T_d}$).

We start by observing that (\ref{BC_MI;JC_IC}) suggests that joint compression can be performed in two steps: ($\mathit{i}$) first, the channel estimate sequence in compressed with required backhaul rate $\frac{1}{T} I(\widetilde {\bf{H}}; \widehat {\bf{H}})$; ($\mathit{ii}$) then, given that the sequence of channel estimates $\widehat {\bf{H}}$ for all coherence blocks is known at both the BS an the CU, the BS uses a different compression strategy for the quantization of ${\bf{Y}}_{d}$ depending on the value of $\widehat {\bf{H}}$\footnote{In practice, the values of $\widehat {\bf{H}}$ can be quantized in order to reduce the number of codebooks.}. Based on this observation, we propose here to adapt the choice of matrix ${\bf{R}}_{d}$ to the current value of $\widehat {\bf{H}}$ for each coherence block. To emphasize this fact, we use the notation ${\bf{R}}_{d}(\widehat {\bf{H}})$. 

\prop For a given adaptive choice ${\bf{R}}_d(\widehat {\bf{H}})$ of the compression covariance matrix on the data signal, the ergodic achievable sum-rate for joint adaptive compression strategy is given as
\begin{eqnarray}
\label{ESR;JAC_IC} R = \frac{T_d}{T} E \left [ \log_2 \det \left( {\bf{I}}_{N_t} + {\frac{P_d}{N_t} \widehat {\bf{H}}^\dagger \left( {\bf{R}}_{d}(\widehat {\bf{H}})  + \sigma_{pe}^2 {\bf{I}}_{N_{r}} \right)^{-1} \widehat {\bf{H}} } \right) \right],
\end{eqnarray} 
where $\widehat {\bf{H}}$ is distributed as in (\ref{CCS+QN}) and $\sigma_{pe}^2$ is given in (\ref{Noise_pe}). Moreover, the quantization noise power $\sigma_p^2$ and the covariance matrices ${\bf{R}}_d(\widehat {\bf{H}})$ must satisfy the backhaul constraint $C_p + C_d = C$, where
\begin{eqnarray}
C_d = \frac{T_d}{T} E \left [ \log_2 \det \left( {\bf{I}}_{N_{r}} + {\bf{R}}_{d}^{-1}(\widehat {\bf{H}}) \left (\frac{P_d}{N_t} \widehat {\bf{H}} \widehat {\bf{H}}^\dagger + \sigma_{pe}^2 {\bf{I}}_{N_{r}} \right) \right) \right] \label{BC;JAC_IC}
\end{eqnarray}
and $C_p$ is defined in (\ref{CBC;SC_IC}).
\begin{IEEEproof}
The ergodic achievable sum-rate follows as for the previous propositions. Moreover, using (\ref{BC_MI;JC_IC}) and following the same steps as in Appendix \ref{Appendix:ProofBC;JC}, we obtain the relationship (\ref{BC;JAC_IC}) between the backhaul capacity and the quantization noise statistics $(\sigma_p^2, {\bf{R}}_d(\widehat {\bf{H}}))$. 
\end{IEEEproof}

We now observe that the optimization of the compression covariance matrices ${\bf{R}}_{d}(\widehat {\bf{H}})$ of the data signal for a given the variance $\sigma_{p}^2$ can be carried out analytically. The problem of maximizing the ergodic achievable sum-rate (\ref{ESR;JAC_IC}) then reduces to a one-dimensional search over $\sigma_p^2$.

\prop \label{S;JAC_IC} Define the eigenvalue decomposition 
\begin{equation}
\frac{P_d}{N_t} \widehat {\bf{H}} \widehat {\bf{H}}^\dagger + \sigma_{pe}^2 {\bf{I}}_{N_{r}} = {\bf{U}}(\widehat {\bf{H}}) \textrm{diag} \left( t_1(\widehat {\bf{H}}), \dots, t_{N_{r}}(\widehat {\bf{H}}) \right) {\bf{U}}^\dagger(\widehat {\bf{H}}). \label{EVD_CM;JAC_IC}
\end{equation} 
The problem of maximizing the ergodic achievable sum-rate (\ref{ESR;JAC_IC}) under the constraint $C = C_p + C_d$, with $C_p$ in (\ref{CBC;SC_IC}) and $C_d$ in (\ref{BC;JAC_IC}), admits the solution ${\bf{R}}_{d}(\widehat {\bf{H}}) = {\bf{U}}(\widehat {\bf{H}}) \textrm{diag} ( \lambda_1(\widehat {\bf{H}}), \dots, \lambda_{N_{r}}(\widehat {\bf{H}}) )^{-1} {\bf{U}}^\dagger(\widehat {\bf{H}})$, where the inverse eigenvalues are given as 
\begin{equation}
\lambda_n^* (\widehat {\bf{H}}) = \left [ \frac{1}{\mu} \left( \frac{1}{\sigma_{pe}^2} - \frac{1}{t_n(\widehat {\bf{H}})} \right) - \frac{1}{\sigma_{pe}^2} \right]^+,   \label{EV;JAC_IC}
\end{equation}
for $n = 1, \dots, N_{r}$; $\sigma_{pe}^2$ is given in (\ref{Noise_pe}); the Lagrange multiplier $\mu^*$ is such that the condition $C = C_p + C_d$, with $C_p$ in (\ref{CBC;SC_IC}) and $C_d$ in (\ref{BC;JAC_IC}), is satisfied with the equality.

\begin{IEEEproof}
The proof follows closely \cite[Theorem 1]{Coso09TWC} and details are available in Appendix \ref{P_S;JAC_IC}.
\end{IEEEproof}
\section{Analysis of ECF : The Multiple Base Stations Case}
\label{Sec:MBC} We now consider the general case with $N_B \ge 1$ BSs. A key aspect that is introduced by the model with multiple BSs is the fact that the signals ${\bf{Y}}_{d,j}$ for $j \in \mathcal{N}_B$ received by the BSs during the data transmission phase are statistically dependent. In fact, they are noisy versions of the same signals transmitted by the MSs.  Therefore, using distributed source coding strategies, the BSs can potentially improve the quality of the descriptions $\widehat {\bf{Y}}_{d,j}$ in (\ref{RDS;CU}) conveyed to the CU over the backhaul links \cite{Sanderovich09TIT}. Note that this is instead not the case for the compression of the channel matrices, since they are assumed to be independent across different BSs\footnote{Strictly speaking, the channel estimates are correlated, due to the correlation of the estimation errors. However, at sufficiently large SNR, this correlation is expected negligible and is hence not further considered here.}. 

A practical way to implement distributed source coding is by means of successive compression \cite{Zhang07TIT}. Accordingly, one defines a permutation $\pi$ of the indices of the BSs. Then, the quantized data signal $\widehat {\bf{Y}}_{d,j}$, for $j \in {\mathcal{N}}_B$, are successively recovered at the CU in the order $\widehat {\bf{Y}}_{d,\pi(1)}, \widehat {\bf{Y}}_{d,\pi(2)}, \dots, \widehat {\bf{Y}}_{d,\pi(N_B)}$. Specifically, when decompressing the signal $\widehat {\bf{Y}}_{d, \pi(j)}$, the CU uses the previously recovered compressed data signals $\widehat {\bf{Y}}_{d, {\mathcal{S}}_j}$, where $\widehat {\bf{Y}}_{d, {\mathcal{S}}_j}$ includes all $\widehat {\bf{Y}}_{d,i}$ with $i \in {\mathcal{S}}_j = \{ \pi (1), \dots, \pi (j-1) \}$. Given the correlation among the received signals, the use of this side information can improve the reproduction quality of the decompressed signals $\widehat {\bf{Y}}_{d,j}$. This has been previously studied in the presence of perfect CSI in \cite{Sanderovich09TIT, Coso09TWC, Park13TVT, Zhou12arXiv}.

In this section, we aim at optimizing the ergodic achievable sum-rate, assuming distributed source coding for the compression of the received data signals, as implemented via successive compression. To this end, similar to \cite{Coso09TWC} \cite{Park13TVT}, we adopt a sequential approach for the optimization of the quantization parameters across the BSs. As in the previous section, we consider compression strategies based on separate, joint, and joint adaptive compression of estimated CSI and received data signal. 

\linespread{1}
\begin{algorithm} [t]
\caption{Greedy algorithm for the multi-BS case} \label{GA}
\begin{algorithmic} [1]
\State Initialize set $\mathcal{S}$ to be an empty set, i.e., $\mathcal{S}_0 = \emptyset$.
\For{$n = 1$ to $N_B$} 
\State Obtain $j^* = \arg \max_{j \in \mathcal{N}_B \backslash \mathcal{S}_{n-1}} R_j^*$ where $R_j^*$ is the optimal value of the problem
\begin{subequations}
\label{P;GA}
\begin{eqnarray}
\label{P_o;GA} \textrm{maximize} && R_j \, \textrm{in} \, (\ref{LB;MBS}) \\
\label{P_c;GA} s.t. && \textrm{backhaul constraint (see Section \ref{Sec:SC;MBC}, C, D)}
\end{eqnarray}
\end{subequations}
\State Update the set $\mathcal{S}_{n} = \mathcal{S}_{n-1} \bigcup \{j^*\}$ and the permutation $\pi^*(n) = j^*$.
\State Assign a solution of (\ref{P;GA}) for $j=j^*$ to the optimal $\sigma_{p, \pi^*(n)}^2$ and ${\bf{R}}_{d,\pi^*(n)}$
\EndFor \\
\Return $\pi^*$, $\{ \sigma_{p,1}^{2}, \dots, \sigma_{p,N_B}^{2} \}$, and $\{{\bf{R}}_{d,1}, \dots, {\bf{R}}_{d,N_B}\}$
\end{algorithmic}
\end{algorithm}
\linespread{2}

\subsection{Problem Definition}
Here we define the optimization problem and the proposed sequential solution. We recall that we need to optimize the compression parameters $(\sigma_{p,j}^2,{\bf{R}}_{d,j})$ for all $j \in {\mathcal{N}}_B$ along with the BS order $\pi$ used for successive compression. Each BS uses the test channel (\ref{CCS+QN}) for the training phase and (\ref{RDS;BS}) for the data phase. Therefore, by the chain rule for the mutual information, given a permutation $\pi$, the ergodic sum-capacity can be written as 
\begin{eqnarray}
\label{OF;MBS} && \frac{1}{T} I \left( {\bf{X}}_d; \widehat {\bf{Y}}_d | \widehat {\bf{H}} \right) = \frac{1}{T} \sum_{j = 1}^{N_B} I \left( {\bf{X}}_d; \widehat {\bf{Y}}_{d, \pi(j)} | \widehat {\bf{H}}, \widehat {\bf{Y}}_{d, {\mathcal{S}}_j} \right). 
\end{eqnarray}
We remark that the rate $\frac{1}{T} I \left( {\bf{X}}_d; \widehat {\bf{Y}}_{d, \pi(j)} | \widehat {\bf{H}}, \widehat {\bf{Y}}_{d, {\mathcal{S}}_j} \right)$ can be interpreted as the contribution of the $j$-th BS to the ergodic sum-capacity. This term can be bounded, similar to the previous sections by overestimating the effect of noise, leading to a lower bound $\frac{1}{T} I \left( {\bf{X}}_d; \widehat {\bf{Y}}_{d, \pi(j)} | \widehat {\bf{H}}, \widehat {\bf{Y}}_{d, {\mathcal{S}}_j} \right) \ge R_j$ (see, Proposition \ref{Prop;MBC_SC} below).

The proposed approach to the optimization of the ergodic achievable sum-rate $\sum_{j=1}^{N_B} R_j$ with respect to the order $\pi$ and the compression parameters $\sigma_{p,j}^2$ and ${\bf{R}}_{d,j}$ for all $j \in {\mathcal{N}}_B$ is summarized in Algorithm \ref{GA}. Specially, we propose a greedy algorithm, whereby at each step, the $j$-th BS is selected that maximizes the contribution $R_j$ of its received signal to the sum-rate. The rate maximization step in (\ref{P;GA}) is discussed in the next section considering separate, joint, or joint adaptive compression building on the analysis in the previous section. Note that the constraint in (\ref{P_c;GA}) depends on the type of compression adopted. Also, we observe that the proposed algorithm can be run at the CU, which only requires knowledge of the statistics of the channels, and that the $j$-th optimal compression parameters ${\sigma_{p,j}^2}$ and ${\bf{R}}_{d,j}$ obtained from Algorithm \ref{GA} can be transmitted to the $j$-th BS by the CU.

\subsection{Separate Compression of Channel and Received Data Signal}
\label{Sec:SC;MBC} In this subsection, we solve the problem (\ref{P;GA}) for a given $j$-th BS assuming separate compression of channel and received data signal. As in Section \ref{Sec:SC;SBC}, we choose ${\bf{R}}_{d,j} = \sigma_{d,j}^2 {\bf{I}}_{N_{r,j}}$ and hence the optimization is over the pair $(\sigma_{p, j}^2, \sigma_{d, j}^2)$. 

\prop \label{Prop;MBC_SC} Let $C_{p,j}$ and $C_{d,j}$ denote respectively the backhaul rates allocated for the transmission of the compressed channel estimates (\ref{CCS+QN}) and of the compressed received signals (\ref{RDS;BS}) on the backhaul link from the $j$-th BS to the CU. For a given a permutation $\pi$, the ergodic achievable sum-rate $R_j$ in (\ref{P_o;GA}) for the $j$-th BS with separate compression strategy is given as
\begin{eqnarray}
\label{LB;MBS} R_j = \frac{T_d}{T} E \left [ \log_2 \det \left( {\bf{I}}_{N_{r, \pi(j)}} + \widehat {\bf{H}}_{\pi(j)} {\bf{R}}_{X|\widehat Y_{{\mathcal{S}}_j}, \widehat H} \widehat {\bf{H}}_{\pi(j)}^\dagger \left( {\bf{R}}_{d,\pi(j)}  + \sigma_{pe, \pi(j)}^2 {\bf{I}}_{N_{r, \pi(j)}} \right)^{-1}  \right) \right],  
\end{eqnarray}
with $\sigma_{pe, j}^2 = \textrm{tr} ( {\bf{R}}_{X|\widehat Y_{{\mathcal{S}}_j}, \widehat H} ) \left( \sigma_{p,j}^2 + \frac{\epsilon_j}{N_t} \right) + 1$, where $\epsilon_{j} = \sum_{i=1}^{N_M} N_{t,i} \sigma_{e_{ji}}^2 $; with $\widehat {\bf{H}}_j$ being distributed as in (\ref{CCS+QN}); and the conditional correlation matrix ${\bf{R}}_{X|\widehat Y_{\mathcal {S}}, \widehat H}$ is defined as 
\begin{eqnarray}
\label{C;X_YSH} {\bf{R}}_{X|\widehat Y_{{\mathcal{S}}_j}, \widehat H} &=& {\bf{R}}_{X} - {\bf{R}}_{X \widehat Y_{{\mathcal{S}}_j}|\widehat H} {\bf{R}}_{\widehat Y_{{\mathcal{S}}_j}|\widehat H}^{-1} {\bf{R}}_{X \widehat Y_{{\mathcal{S}}_j}|\widehat H}^\dagger \\
\nonumber &=& \frac{P_d}{N_t} {\bf{I}}_{N_t} - \left( \frac{P_d}{N_t} \right)^2 \widehat {\bf{H}}_{{\mathcal{S}}_j}^\dagger \left( \frac{P_d}{N_t} \left( \widehat {\bf{H}}_{{\mathcal{S}}_j} \widehat {\bf{H}}_{{\mathcal{S}}_j}^\dagger + N_t {\bf{R}}_{p, {\mathcal{S}}_j} + {\bf{R}}_{\epsilon,\mathcal{S}_j} \right) + {\bf{R}}_{d,{\mathcal{S}}_j} + {\bf{I}}_{N_{r,{\mathcal{S}}_j}} \right)^{-1} \widehat {\bf{H}}_{{\mathcal{S}}_j},
\end{eqnarray}
with ${\bf{R}}_{d, {\mathcal{S}}_j}$, ${\bf{R}}_{p, {\mathcal{S}}_j}$ and ${\bf{R}}_{\epsilon,\mathcal{S}_j}$ being block diagonal matrices with main diagonals given by $[{\bf{R}}_{d,\pi(1)}, \dots, {\bf{R}}_{d,\pi(j-1)}]$, $[\sigma_{p,\pi(1)}^2 {\bf{I}}_{N_{r, \pi(1)}}, \dots, \sigma_{p,\pi(j-1)}^2 {\bf{I}}_{N_{r, \pi(j-1)}}]$ and $[\epsilon_{\pi(1)}{\bf{I}}_{N_{r, \pi(1)}}, \dots, \epsilon_{\pi(j-1)}{\bf{I}}_{N_{r, \pi(j-1)}}]$, respectively. Moreover, the quantization noise powers $(\sigma_{p,j}^2, \sigma_{d,j}^2)$ for the $j$-th BS must satisfy the backhaul constraint $C_{p,j} + C_{d,j} = C_j$ in (\ref{P_c;GA}), where 
\begin{subequations}
\label{BR;ECF_Sep_MBC}
\begin{eqnarray}
\label{BR;ECF_Sep_MBC_DS} C_{p,j} &=& \frac{N_{r,j}}{T} \log_2 \left( \frac{ \prod_{i=1}^{N_M} \left( \sigma_{\widetilde h_{ji}}^2\right)^{N_{t,i}}}{(\sigma_{p,j}^2)^{N_t}} \right) \\
\label{BR;ECF_Sep_MBC_TS}  \textrm{and} \hspace{0.5cm}  {C_{d,j}} &=& \frac{T_d}{T} \left( \log_2 \det \left( {\bf{I}}_{N_{r,j}} + \frac{E \left[ \widehat {\bf{H}}_j {\bf{R}}_{X|\widehat Y_{\mathcal {S}_j}, \widehat H} \widehat {\bf{H}}_j^\dagger + \left( \sigma_{p,j}^2 + \frac{\epsilon_j}{N_t} \right) {\bf{R}}_{X|\widehat Y_{\mathcal {S}_j}, \widehat H} \right] + {\bf{I}}_{N_{r,j}}}{\sigma_{d,j}^2} \right) \right),
\end{eqnarray}
\end{subequations}
with $\sigma_{\widetilde h_{ji}}^2$ being given in (\ref{MMSE}).
\begin{IEEEproof}
The ergodic achievable sum-rate $R_j$ is evaluated as in Lemma \ref{lemma;Pre_CFE}. As for the backhaul constraint, the only difference with respect to Section \ref{Sec:SC;SBC} is the presence of the side information $(\widehat {\bf{Y}}_{\mathcal{S}_j}, \widehat {\bf{H}}_{\mathcal{S}_j})$ at the CU. Since the channel and side information are independent, the relationships (\ref{CBC;SC_IC})-(\ref{CV_CQN}) between the CSI quantization error $\sigma_{p,j}^2$ and $C_{p,j}$ are unchanged, and hence the backhaul rate used for transmitting the estimated CSI can be written as (\ref{BR;ECF_Sep_MBC_DS}). Instead, using the well-known Wyner-Ziv theorem (see, e.g., \cite[Section 11.3]{GamalBook}), the rate needed to compress the data received signal ${\bf{Y}}_{d,j}$ given the side information $( \widehat {\bf{Y}}_{d,\mathcal{S}_j}, \widehat {\bf{H}}_{\mathcal{S}_j})$ available at the CU is given by (cf. (\ref{DBC;SC_IC}))
\begin{eqnarray}
\label{DBC;SC_DC} \nonumber {C_{d,j}} &=& \frac{1}{T} I \left( {\bf{Y}}_{d,j};\widehat {\bf{Y}}_{d,j} | \widehat {\bf{Y}}_{d, \mathcal{S}_j}, \widehat {\bf{H}}_{\mathcal{S}_j} \right) \\ &\le& \frac{T_d}{T} \left( \log_2 \det \left( {\bf{I}}_{N_{r,j}} + \frac{E \left[ \widehat {\bf{H}}_j {\bf{R}}_{X|\widehat Y_{\mathcal {S}_j}, \widehat H} \widehat {\bf{H}}_j^\dagger + \left( \sigma_{p,j}^2 + \frac{\epsilon_j}{N_t} \right) {\bf{R}}_{X|\widehat Y_{\mathcal {S}_j}, \widehat H} \right] + {\bf{I}}_{N_{r,j}}}{\sigma_{d,j}^2} \right) \right).
\end{eqnarray}
\end{IEEEproof}
Note that for $j=1$, the rate (\ref{LB;MBS}) and backhaul rate (\ref{DBC;SC_DC}) equal (\ref{EAR;SC_IC}) and (\ref{BR;ECF_Sep_TS}), respectively. Moreover, the optimization of (\ref{LB;MBS}) requires a one-dimensional search over $C_{p,j}$ or $C_{d,j}$ as for the single BS case in Section \ref{Sec:SC;SBC}.

\rem As Remark \ref{rem:SBC}, with Rayleigh fading (i.e., $K=0$), we can calculate the quantization error variance $\sigma_{d,j}^2$ by solving (\ref{DBC;SC_DC}) as
\begin{equation}
\sigma_{d,j}^2 = \frac{ P_d - \frac{P_d^2}{N_t} \left( \sum_{k \in \mathcal{S}_j} \frac{N_{r,k} \left( 1 - \sigma_{p,k}^2 - \frac{\epsilon_k}{N_t} \right)}{1 + P_d + \sigma_{d,k}^2} \right) + 1}{2^{T C_{d,j} / N_{r,j} T_d} - 1}. \vspace{-0.6cm}
\end{equation} 
$\hspace{17.8cm} \blacksquare$
 
\subsection{Joint Compression of Channel and Received Data Signal}
\label{Sec:JC;MBC} We now tackle problem (\ref{P;GA}) assuming joint compression of channel and received data signal. As in Section \ref{Sec:JC;SBC}, we assume an uncorrelated compression covariance ${\bf{R}}_{d,j} = \sigma_{d,j}^2 {\bf{I}}_{N_{r,j}}$ in the test channel (\ref{RDS;BS}). 

\prop For a given a permutation $\pi$, the ergodic achievable sum-rate $R_j$ for the $j$-th BS with joint compression strategy is given by (\ref{LB;MBS}). Moreover, the quantization noise powers $(\sigma_{p,j}^2, \sigma_{d,j}^2)$ for the $j$-th BS must satisfy the backhaul constraint $C_{p,j} + C_{d,j} = C_j$ in (\ref{P_c;GA}), where 
\begin{eqnarray}
C_{d,j} = \frac{T_d}{T} \left( E \left [ \log_2 \det \left( \widehat {\bf{H}}_j  {\bf{R}}_{X|\widehat Y_{\mathcal {S}_j}, \widehat H} \widehat {\bf{H}}_j^\dagger + \left( \sigma_{pe, j}^2 + \sigma_{d, j}^2 \right) {\bf{I}}_{N_{r,j}} \right)  \right] - N_{r,j} \log_2 \left( \sigma_{d,j}^2 \right)  \right), \label{CV_BC;JC_DC}
\end{eqnarray}
and $C_{p,j}$ is defined in (\ref{BR;ECF_Sep_MBC_DS}), with $\sigma_{pe, j}^2 = \textrm{tr} ( {\bf{R}}_{X|\widehat Y_{{\mathcal{S}}_j}, \widehat H} ) \left( \sigma_{p,j}^2 + \frac{\epsilon_j}{N_t} \right) + 1$ and ${\bf{R}}_{X|\widehat Y_{\mathcal {S}}, \widehat H}$ being defined in (\ref{C;X_YSH}). 
\begin{IEEEproof}
Following similar considerations as above and as in Section \ref{Sec:JC;SBC}, given side information $\widehat {\bf{Y}}_{d, \mathcal{S}_j}$ and $\widehat {\bf{H}}_{\mathcal{S}_j}$, the rate required on the backhaul with joint compression of channel and received data signal is 
\begin{eqnarray}
C_{j} = \frac{1}{T} I \left( {\bf{Y}}_{d,j},\widetilde {\bf{H}}_j;\widehat {\bf{Y}}_{d,j},\widehat {\bf{H}}_j | \widehat {\bf{Y}}_{d, \mathcal{S}_j}, \widehat {\bf{H}}_{\mathcal{S}_j} \right) = \frac{1}{T} \left( I \left( \widetilde {\bf{H}}_j; \widehat {\bf{H}}_j \right) + I \left( {\bf{Y}}_{d,j}; \widehat {\bf{Y}}_{d,j} | \widehat {\bf{H}}_j , \widehat {\bf{Y}}_{d, \mathcal{S}_j}, \widehat {\bf{H}}_{\mathcal{S}_j} \right) \right), \label{BC;JC_DC}
\end{eqnarray}
where the second equality can be shown similar to the derivations in Appendix \ref{Appendix:ProofBC;JC}. From the maximum entropy theorem, the equality (\ref{BC;JC_DC}) implies the constraint $C_j = C_{p,j} + C_{d,j}$ with $C_{p,j}$ in (\ref{BR;ECF_Sep_MBC_DS}) and $C_{d,j}$ in (\ref{CV_BC;JC_DC}).
\end{IEEEproof}
Note that for $N_B=1$, (\ref{CV_BC;JC_DC}) reduces to (\ref{BC;JC_IC}). 
Furthermore, maximization of (\ref{LB;MBS}) requires a search over the space $(\sigma_{p,j}^2, \sigma_{d,j}^2)$ as for the single BS case in Section \ref{Sec:JC;SBC}.
%
\subsection{Joint Adaptive Compression of Channel and Received Data Signal}
\label{Sec:JAC;MBC} Considering joint adaptive compression, the backhaul constraint is still given by (\ref{BC;JC_DC}), but now we consider the quantization noise ${\bf{Q}}_{d,j}$ to have a covariance matrix ${\bf{R}}_{d,j}$ that is allowed to depend on the channel estimate $\widehat {\bf{H}}_j$ and on the estimates $\widehat {\bf{H}}_{\mathcal{S}_j}$ of the previously selected BSs. 

\prop  For a given a permutation $\pi$, the ergodic achievable sum-rate $R_j$ for the $j$-th BS with joint adaptive compression strategy is given as (\ref{LB;MBS}) with ${\bf{R}}_{d,j}(\widehat {\bf{H}}_{\mathcal{S}_j \cup \{j\}})$ in lieu of ${\bf{R}}_{d,j}$. Moreover, the quantization noise power $\sigma_{p,j}^2$ and the covariance matrices ${\bf{R}}_{d,j}(\widehat {\bf{H}}_{\mathcal{S}_j \cup \{j\}})$ for the $j$-th BS must satisfy the backhaul constraint $C_{p,j} + C_{d,j} = C_j$ in (\ref{P_c;GA}), where 
\begin{eqnarray}
C_{d,j} = \frac{T_d}{T} E \left [ \log_2 \det \left( {\bf{I}}_{N_{r,j}} + {\bf{R}}_{Y_j|\widehat Y_{\mathcal {S}_j}, \widehat H} {\bf{R}}_{d,j}^{-1}(\widehat {\bf{H}}_{\mathcal{S}_j \cup \{j\}}) \right) \right], \label{BC;JAC_MBC}
\end{eqnarray}
as a function of ${\bf{R}}_{d,j}(\widehat {\bf{H}}_{\mathcal{S}_j \cup \{j\}})$, $C_{p,j}$ is defined in (\ref{BR;ECF_Sep_MBC_DS}) and we have
\begin{eqnarray}
{\bf{R}}_{Y_j|\widehat Y_{\mathcal {S}_j}, \widehat H} = \widehat {\bf{H}}_j {\bf{R}}_{X|\widehat Y_{\mathcal {S}_j}, \widehat H} \widehat {\bf{H}}_j^\dagger + \sigma_{pe, j}^2 {\bf{I}}_{N_{r,j}}, \label{C;Y|Y,H}
\end{eqnarray}
with $\sigma_{pe, j}^2 = \textrm{tr} ( {\bf{R}}_{X|\widehat Y_{{\mathcal{S}}_j}, \widehat H} ) \left( \sigma_{p,j}^2 + \frac{\epsilon_j}{N_t} \right) + 1$.
\begin{IEEEproof}
Using (\ref{BC;JC_DC}) and following similar steps as in Appendix \ref{Appendix:ProofBC;JC}, we obtain the relationship $C_j = C_{p,j} + C_{d,j}$, with $C_{p,j}$ in (\ref{BR;ECF_Sep_MBC_DS}) and $C_{d,j}$ in (\ref{BC;JAC_MBC}), between the backhaul capacity and the quantization noise statistics $(\sigma_{p,j}^2, {\bf{R}}_{d,j}(\widehat {\bf{H}}_{\mathcal{S}_j \cup \{j\}}))$. 
\end{IEEEproof}
As in Section \ref{Sec:JAC;SBC}, we can now solve problem (\ref{P;GA}) with respect to the compression covariance matrix ${\bf{R}}_{d,j} ( \widehat {\bf{H}}_{\mathcal{S}_j \cup \{j\}} )$, as reported in the proposition below.

\prop \label{S;JAC_MBC} Define the eigenvalue decomposition 
\begin{eqnarray}
{\bf{R}}_{Y_j|\widehat Y_{\mathcal {S}}, \widehat H} = {\bf{U}}_j(\widehat {\bf{H}}_{\mathcal{S}_j \cup \{j\}}) \textrm{diag} ( t_1(\widehat {\bf{H}}_{\mathcal{S}_j \cup \{j\}}), \dots, t_{N_{r,j}}(\widehat {\bf{H}}_{\mathcal{S}_j \cup \{j\}}) ) {\bf{U}}_j^\dagger(\widehat {\bf{H}}_{\mathcal{S}_j \cup \{j\}}).
\end{eqnarray}
The problem of maximizing the ergodic achievable sum-rate (\ref{LB;MBS}) under the constraint $C_j = C_{p,j} + C_{d,j}$, with $C_{p,j}$ in (\ref{BR;ECF_Sep_MBC_DS}) and $C_{d,j}$ in (\ref{BC;JAC_MBC}), admits the solution ${\bf{R}}_{d,j}(\widehat {\bf{H}}_{\mathcal{S}_j \cup \{j\}}) = {\bf{U}}_j(\widehat {\bf{H}}_{\mathcal{S}_j \cup \{j\}}) \textrm{diag} ( \lambda_1(\widehat {\bf{H}}_{\mathcal{S}_j \cup \{j\}}), \dots, \lambda_{N_{r,j}}(\widehat {\bf{H}}_{\mathcal{S}_j \cup \{j\}}) )^{-1}$ ${\bf{U}}_j^\dagger(\widehat {\bf{H}}_{\mathcal{S}_j \cup \{j\}})$, where the inverse eigenvalues are given as 
\begin{equation}
\lambda_n^* (\widehat {\bf{H}}_{\mathcal{S}_j \cup \{j\}}) = \left [ \frac{1}{\mu_j} \left( \frac{1}{\sigma_{pe, j}^2 } - \frac{1}{t_n(\widehat {\bf{H}}_{\mathcal{S}_j \cup \{j\}})} \right) - \frac{1}{\sigma_{pe, j}^2 } \right]^+, \label{EV;JAC_MBC}
\end{equation}
for all $n=1, \dots, N_{r,j}$; $\sigma_{pe, j}^2$ is given in (\ref{LB;MBS}); the Lagrange multiplier $\mu_j^*$ is such that the condition $C_j = C_{p,j} + C_{d,j}$, with $C_{p,j}$ in (\ref{BR;ECF_Sep_MBC_DS}) and $C_{d,j}$ in (\ref{BC;JAC_MBC}), is satisfied with equality.
\begin{IEEEproof}
The proof follows in a similar fashion as Proposition \ref{S;JAC_IC} and is not detailed here. 
\end{IEEEproof}
\section{Semi-Coherent Processing}
\label{Sec:NCnSCP}
In Section \ref{subSec:NCP_Pre}, we have discussed the reference non-coherent strategy, whereby no pilots are transmitted. In the following sections, we have instead elaborated on the CFE and ECF schemes that transfer pilot information or CSI from the BS to the CU over the backhaul links. Here, we propose a novel ``semi-coherent" scheme that, similar to non-coherent processing, operates without transmitting CSI or pilot information to the CU, although pilot signals are transmitted by the MSs as in the CFE and ECF schemes. With the proposed semi-coherent approach, each BS estimates the CSI, performs local equalization and compresses the equalized signal. The CU then performs joint decoding using a mismatched decoding metric \cite{Weingarten04TIT}. Since the analysis of this scheme is an open problem in the presence of multiple MSs, even with a single BS and ideal backhaul, we focus here on a single MS and single BS for simplicity of analysis. This case is expected to provide insight that carry over to more general scenarios.

The MS operates as described in Section \ref{Sec:SM}, while the BS estimates the CSI as in (\ref{MMSE}) and then equalizes the received data signal. Recall that the latter is given in (\ref{RS_d;BS}) and hence can be written as ${\bf{Y}}_d = \widetilde{\bf{H}} {\bf{X}}_d + \widetilde {\bf{Z}}_d$, where the estimated channel $\widetilde {\bf{H}}$ is defined in (\ref{MMSE}) and the equivalent noise is given as $\widetilde {\bf{Z}}_d = {\bf{E}} {\bf{X}}_d + {\bf{Z}}_d$ with channel estimation error ${\bf{E}}$ in (\ref{ECS+EE}). 

The BS performs MMSE equalization\footnote{Other types of  linear equalization could be considered as well following the same steps.} of the data signal based on the channel estimate $\widetilde {\bf{H}}$. Accordingly, we can write the equalized signal as
\begin{eqnarray}
\label{Equalized_DS;SCP} {\bf{G}} {\bf{Y}}_d = {\bf{X}}_d + \left( {\bf{G}} \widetilde {\bf{H}} - {\bf{I}}_{N_t} \right) {\bf{X}}_d + {\bf{G}} \widetilde {\bf{Z}}_d,
\end{eqnarray}
where the equalizing matrix ${\bf{G}}$ is given as ${\bf{G}} = (\widetilde {\bf{H}}^\dagger \widetilde {\bf{H}} + (\sigma_e^2 + \frac{N_t}{P_d}) {\bf{I}}_{N_t} )^{-1} \widetilde {\bf{H}}^\dagger$. The equalized data signal (\ref{Equalized_DS;SCP}) is compressed by the BS and forwarded to the CU on the backhaul link. The compressed equalized data signal $\widehat {\bf{X}}_d$ is obtained as 
\begin{eqnarray}
\label{Comp_DS;SCP} \nonumber \widehat {\bf{X}}_d &=& {\bf{X}}_d + \left( {\bf{G}} \widetilde {\bf{H}} - {\bf{I}}_{N_t} \right) {\bf{X}}_d + {\bf{G}} \widetilde {\bf{Z}}_d + {\bf{Q}}_d \\
&=& {\bf{X}}_d + \widehat {\bf{Z}}_d,
\end{eqnarray}
where the quantization noise matrix ${\bf{Q}}_d$ has i.i.d. ${\mathcal{CN}} (0, \sigma_d^2)$ entries, and the effective noise $\widehat {\bf{Z}}_d$, conditioned on the channel estimate $\widetilde {\bf{H}}$, has covariance matrix 
\begin{equation}
\label{CovN;SCP} {\bf{R}}_{\widehat Z|\widetilde H} = \frac{P_d}{N_t} ({\bf{G}}\widetilde{\bf{H}} - {\bf{I}}_{N_t} )({\bf{G}}\widetilde{\bf{H}} - {\bf{I}}_{N_t} )^\dagger + (\frac{P_d}{N_t} \sigma_e^2 + 1) {\bf{G}} {\bf{G}}^\dagger + \sigma_d^2 {\bf{I}}_{N_t}. 
\end{equation}

From the compressed signal in (\ref{Comp_DS;SCP}), the CU performs decoding by choosing the codeword $({\bf{X}}_{d,1}, \dots, {\bf{X}}_{d,n})$ in the codebook, where $n$ is the number of coherence blocks on which coding takes place. Given the lack of CSI at the receiver, investigating the performance of the optimal, maximum likelihood, decoder is not an easy task. To tackle this issue, we assume that the receiver employs the mismatched nearest neighbor metric 
\begin{eqnarray}
\label{neighbor_metric} \sum_{k=1}^n \gamma_k \parallel \widehat {\bf{X}}_{d,k} - {\bf{X}}_{d,k} \parallel^2.
\end{eqnarray}
In (\ref{neighbor_metric}), the weighting factors $\gamma_k$ are known to the CU, as further discussed below, and hence the metric (\ref{neighbor_metric}) can be computed at the CU even in the absence of CSI. It is also noted that the metric (\ref{neighbor_metric}) is generally mismatched to the actual signal model (\ref{Comp_DS;SCP}), since in (\ref{CovN;SCP}) the noise covariance ${\bf{R}}_{Z|\widetilde H}$ is not a multiple of the identity matrix and depends on the channel estimate $\widetilde {\bf{H}}$, which is not known at the CU. 

We first consider the case in which an equal weighting factor is used in (\ref{neighbor_metric}) for all coherence blocks, i.e., $\gamma_k = \gamma$ for all $k$, and hence the metric (\ref{neighbor_metric}) reduces to $\sum_k \gamma ||\widehat {\bf{X}}_{d,k} - {\bf{X}}_{d,k}||^2$. An ergodic rate achievable with scheme is derived next.

\lemma \label{lemma;Const_Semi-CP} An ergodic achievable rate with semi-coherent processing and constant weights $\gamma_k = \gamma$ in (\ref{neighbor_metric}) is given by 
\begin{eqnarray}
\label{ER;SemiwoCSI} R = \frac{T_d}{T} \sup_{\gamma > 0} \left\{N_t \log_2 \left(1 + \gamma \frac{P_d}{N_t} \right) + \gamma P_d \left(1 + \gamma \frac{P_d}{N_t} \right)^{-1} - \gamma^2 \frac{P_d}{N_t} \left(1+ \gamma \frac{P_d}{N_t} \right)^{-1} E\left[ \textrm{tr} \left({\bf{R}}_{\widehat Z|\widetilde H}\right) \right] \right\},
\end{eqnarray}
where ${\bf{R}}_{\widehat Z|\widetilde H}$ is given in (\ref{CovN;SCP}) and we have $\sigma_d^2 = \frac{P_d + 1}{2^{TC/ (N_rT_d)} -1}$. The expression in (\ref{ER;SemiwoCSI}) is taken with respect to $\widetilde {\bf{H}}$.
\begin{IEEEproof}
The equation (\ref{ER;SemiwoCSI}) follows immediately from \cite[Eq. (19)]{Weingarten04TIT}. 
\end{IEEEproof}

Next, we briefly consider also the possibility to choose the weighting factors $\gamma_k$ in the decoding metric (\ref{neighbor_metric}) as a function of a one-bit per-coherence block CSI sent on the backhaul from BS to CU. Specifically, we fix a threshold $\omega \ge 0$ on the CSI. Then, we choose the weighting coefficient $\gamma_k$ to be small, $\gamma_k = \gamma_b$, when the CSI is of poor quality, i.e., $||\widetilde {\bf{H}}|| < \omega$, and to be large, $\gamma_k = \gamma_g$, when the CSI is of good quality, i.e., $||\widetilde {\bf{H}}|| \ge \omega$. The idea is that coherence blocks with poor CSI should be weighted less. Note that the one-bit CSI message on the backhaul requires the condition $C > 1/T$ to be satisfied.

\prop An ergodic rate achievable with semi-coherent processing and selective weights is given as 
\begin{eqnarray}
\label{ER;SemiwCSI} R = \frac{T_d}{T} \sup_{\gamma_b, \gamma_g, \omega > 0} \left\{ E \left[ N_t \log_2 \left(1 + \Gamma \frac{P_d}{N_t} \right) + \Gamma P_d \left(1 + \Gamma \frac{P_d}{N_t} \right)^{-1} - \Gamma^2 \frac{P_d}{N_t} \left(1+ \Gamma \frac{P_d}{N_t} \right)^{-1} \textrm{tr} \left({\bf{R}}_{\widehat Z|\widetilde H}\right) \right] \right\}, 
\end{eqnarray}
where ${\bf{R}}_{\widehat Z|\widetilde H}$ is given in (\ref{CovN;SCP}) and we have $\sigma_d^2= \frac{P_d + 1}{2^{(TC-1)/ (N_rT_d)} -1}$ The expectation in (\ref{ER;SemiwCSI}) is taken with respect to $\widetilde {\bf{H}}$ and to the random variable $\Gamma$, which is defined as $\Gamma = \gamma_b$ with probability $\textrm{Pr} [||\widetilde {\bf{H}}|| < \omega ]$ and $\Gamma = \gamma_g$ with probability $\textrm{Pr} [ ||\widetilde {\bf{H}}|| \ge \omega ]$. 
\begin{IEEEproof}
The equation (\ref{ER;SemiwCSI}) follows again directly from \cite[Eq. (19)]{Weingarten04TIT}.
\end{IEEEproof}

\section{Numerical Results}
\label{Sec:Numerical Results}
In this section, we evaluate the performance of the proposed compression strategies for the uplink of a multi-cell system. Throughout, we assume that every MS is subject to the same power constraint $P$ and that each BS has the same backhaul capacity $C$, that is $P_i = P$ for $i \in \mathcal{N}_M$ and $C_j = C$ for $j \in \mathcal{N}_B$. Moreover, we set $\bar {\bf{H}}_j = {\bf{1}}_{N_{r,j} \times N_t}$. We optimize over the power allocation $(P_p, P_d)$ and we set $T_p = N_t$ (except for the non-coherent scheme where $T_p=0$), which was shown to be optimal in \cite{Hassibi03TIT} for a point-to-point link with no backhaul limitation.

\begin{figure}[t]
\centering
\includegraphics[width=13.5cm]{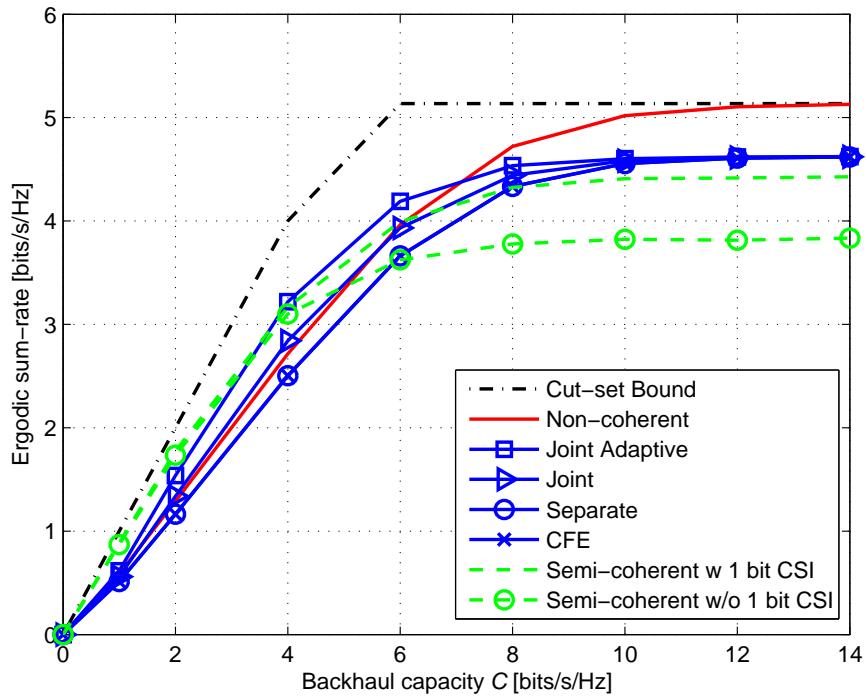}
\caption{Ergodic achievable sum-rate vs. backhaul capacity ($N_B = N_M = 1$, $N_t = N_r=1$, $P=20dB$, $T$ = 10, and $K=0$).}
\label{Fig:SISOBSvsBC}
\end{figure}

\begin{figure}[t]
\centering
\includegraphics[width=13.5cm]{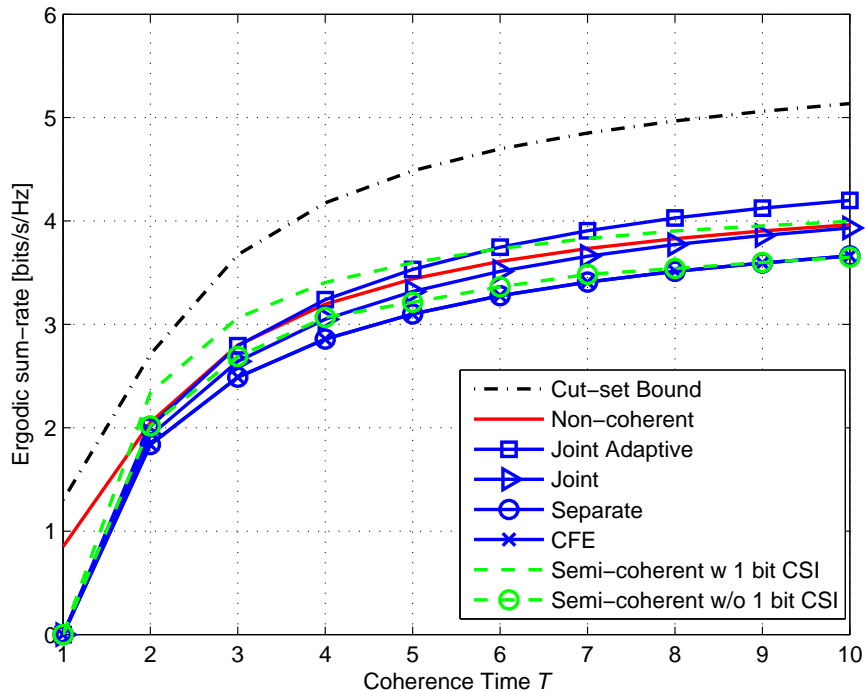}
\caption{Ergodic achievable sum-rate vs. coherence time ($N_B = N_M = 1$, $N_t = N_r=1$, $C$ = 6, $P=20dB$, and $K=0$).}
\label{Fig:SISOBSvsCT}
\end{figure}

We start by considering case of a single MS and a single BS, namely $N_B=1$ and $N_M=1$ and consider the performance of the ECF schemes, of CFE and of non-coherent and semi-coherent processing. For the latter, we focus on the semi-coherent scheme with one-bit CSI and without one-bit CSI. Fig. \ref{Fig:SISOBSvsBC} and Fig. \ref{Fig:SISOBSvsCT} show the ergodic achievable sum-rate for all the mentioned schemes as function of the backhaul capacity $C$ and coherence time $T$\footnote{Consider a multicarrier system. The coherence bandwidth can be approximated as $1/(50 \sigma_\tau)$, where $\sigma_\tau$ is the delay spread \cite{Sklar97COMMMAG}. Therefore, by imposing $1/(50 \sigma_\tau)=T \Delta f$, where $\Delta f$ is the subcarrier spacing, one can find that a delay spread equal to $\sigma_\tau= 1/(50 T \Delta f)$ causes a coherent block equal to $T$ channel uses. For instance, with $\Delta f = 15kHz$, as for LTE systems, we get that $T=1$ corresponds to $\sigma_\tau = 13 \mu s$.}, respectively. For reference, in both figures, we also show the upper bound obtained by standard cut-set arguments, namely $\min(C, R_{nc})$, where $R_{nc}$ is the non-coherent capacity of the MS-BS channel \cite{Marzetta99TIT}. In Fig. \ref{Fig:SISOBSvsBC}, we set $N_t=N_r=1$, power $P = 20dB$, coherence time $T=10$ and consider Rayleigh fading channel, i.e., $K = 0$. At low backhaul capacity $C$ (here, $C<4$), it is seen that the semi-coherent strategy is to be preferred due to its ability to devote the limited backhaul resources to convey only information about the data block to the CU. 
Note that the semi-coherent scheme with one-bit CSI outperforms the case with no CSI unless the backhaul capacity $C$ is smaller or very close to $1/T$ (i.e., the overhead for the one-bit CSI on the backhaul). Conversely, for sufficiently large backhaul capacities (here, $C>7$), the non-coherent approach turns out to be advantageous. This is because, when the compression noise is negligible, the achievable rate is upper bounded by the non-coherent capacity\footnote{In a non-coherent information-theoretic set-up, the optimization of the transmit signals allows, as a special case, the selection of a pilot-based transmission in which all codewords contain the same training sequence.} (see, e.g., \cite{Marzetta99TIT}). Instead, for intermediate backhaul values, ECF and CFE schemes are the preferred choice. Concerning the comparison between ECF and CFE, Fig. \ref{Fig:SISOBSvsBC} demonstrates that the ECF strategy is advantageous. In particular, for the scenario at hand, CFE performs as ECF with separate compression as discussed in Section \ref{Sec:SC;SBC}. However, progressively more complex ECF schemes have better performance, with the joint adaptive strategy outperforming the joint approach and the separate strategy. Finally, we note that the gains obtained by more complex ECF compression strategies are especially pronounced in the region of interest of moderate backhaul capacity, in which the backhaul capacity is at a premium and should be used efficiently. 

The effect of an increase of the coherence time on the ergodic achievable sum-rate is instead investigated with $N_t=N_r=1$, backhaul capacity $C=6$, power $P = 20dB$, and Rayleigh fading in Fig. \ref{Fig:SISOBSvsCT}. The figure illustrates that the non-coherent strategy is clearly advantageous over the other schemes for $T=1$ given that it operates without transmitting any pilot signal. Moreover, ECF with Joint adaptive compression is especially advantageous for large coherence time due to the increased relevance of an efficient compression of the data signal when $T_d \gg T_p$.  

\begin{figure}[t]
\centering
\includegraphics[width=13.5cm]{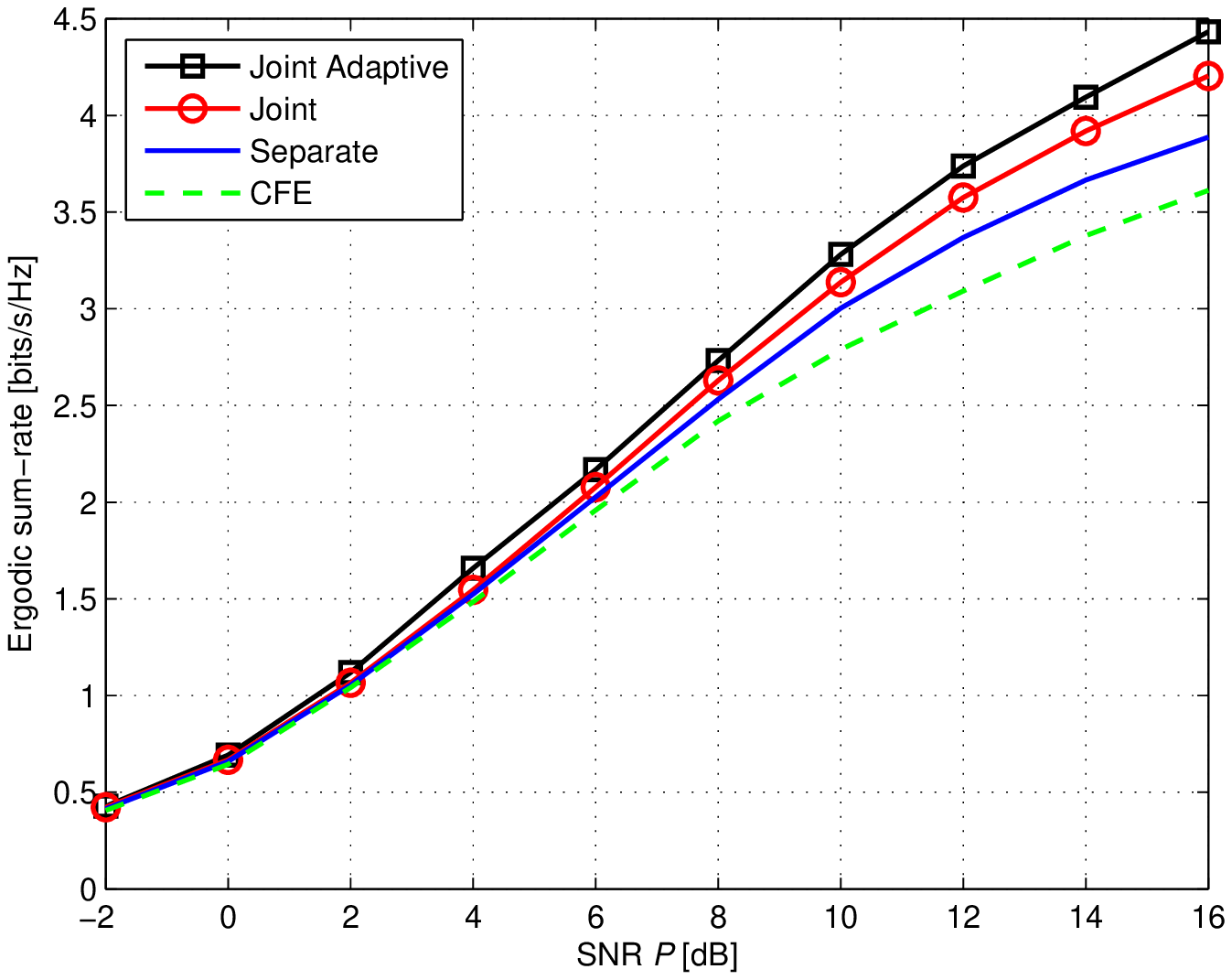}
\caption{Ergodic achievable sum-rate vs. SNR (dB) ($N_B = N_M = 2$, $N_t = N_r=4$, $C=6$, $T= 10$, $\alpha_{ji}=1$, and $K=0$).}
\label{Fig:MultipleBSvsSNR}
\end{figure}
\begin{figure}[h!]
\centering
\includegraphics[width=13.5cm]{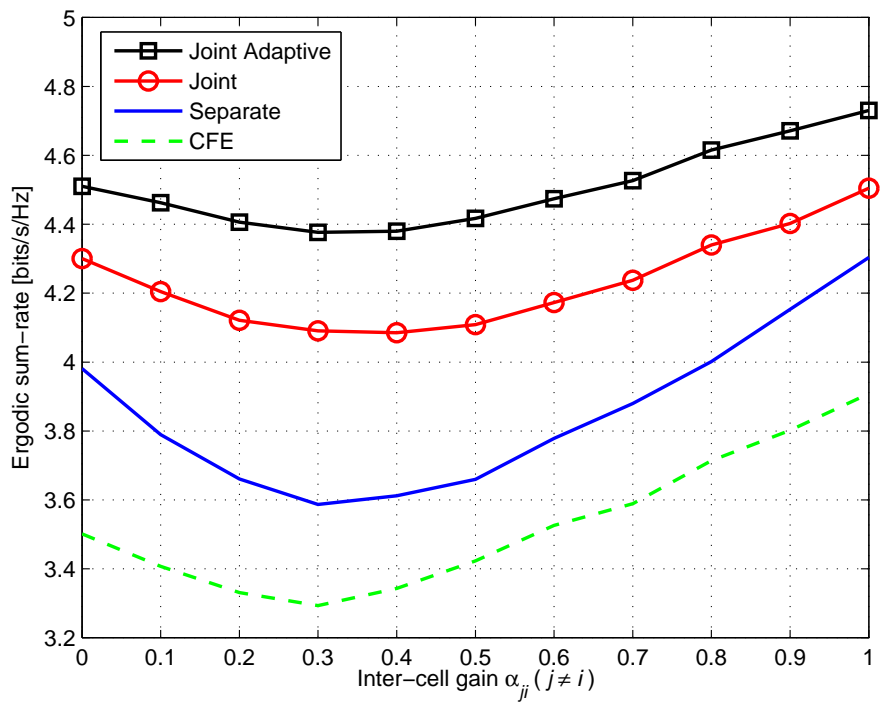}
\caption{Ergodic achievable sum-rate vs. inter-cell gain $\alpha_{ji}$  ($N_B = N_M = 2$, $N_t = N_r=4$, $C=6$, $P = 20dB$, $T=10$ and $K=0$).}
\label{Fig:MultipleBSvsAlpha}
\end{figure}

\begin{figure}[t]
\centering
\includegraphics[width=13.5cm]{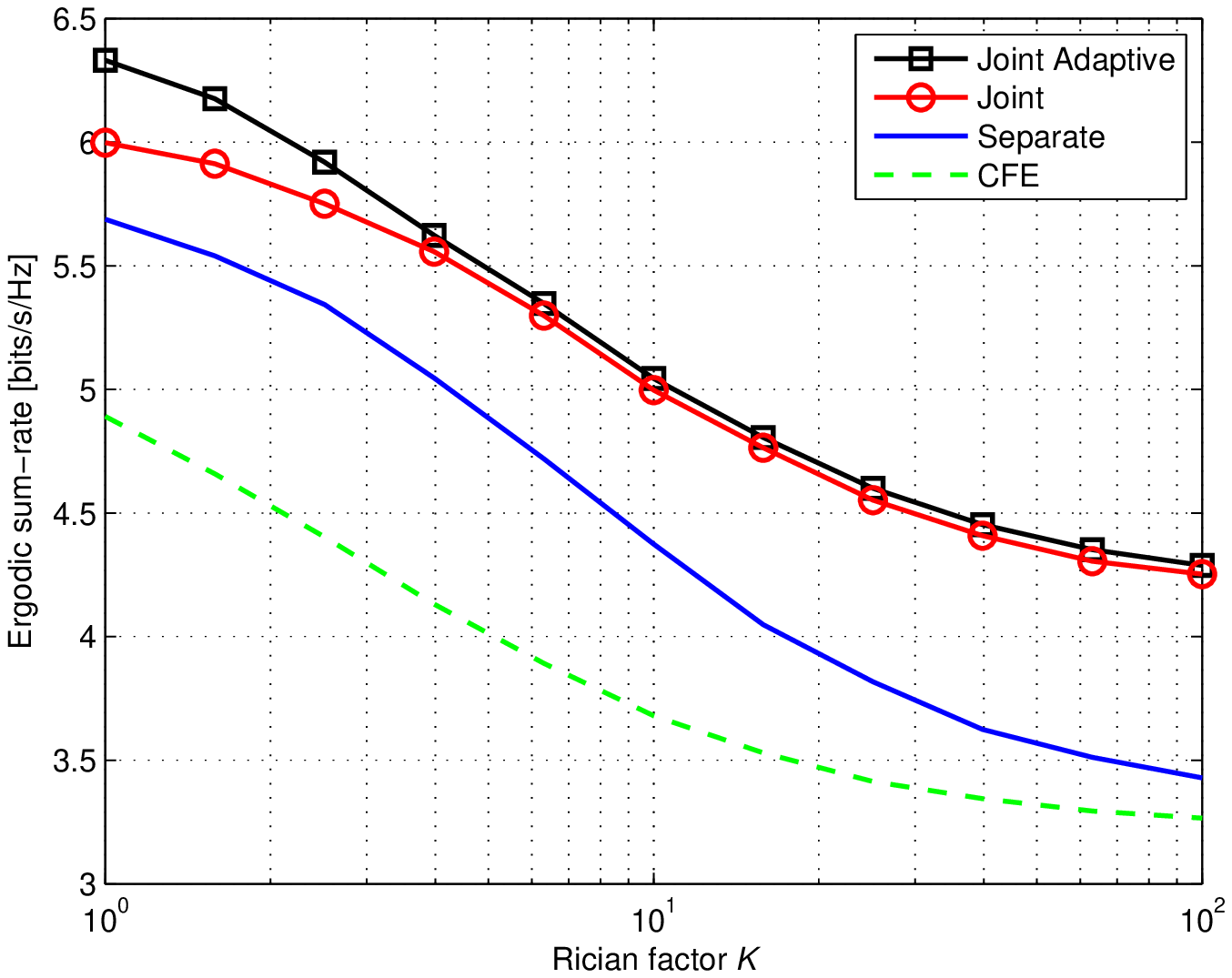}
\caption{Ergodic achievable sum-rate vs. Rician factor $K$ ($N_B = N_M = 2$, $N_t = N_r=4$, $C=6$, $P = 20dB$, $T=20$ and $\alpha_{ji}=1$).}
\label{Fig:MultipleBSvsRF}
\vspace{-0.3cm}
\end{figure}

We now turn to consider a multiple BSs and multiple MSs scenario with $N_B=N_M=2$, $N_t=N_r=4$ and focus on the comparison among the different proposed ECF schemes and CFE\footnote{With multiple BSs and MSs, evaluating the non-coherent capacities, and thus also the cut-set bound is an open problem. Moreover, the evaluation of the performance of semi-coherent strategies is left for future work.}. The performance comparison among the proposed ECF schemes discussed above is confirmed by the results reported in Fig. \ref{Fig:MultipleBSvsSNR}, \ref{Fig:MultipleBSvsAlpha} and \ref{Fig:MultipleBSvsRF}. Fig. \ref{Fig:MultipleBSvsSNR} shows the ergodic achievable sum-rate of the three compression methods versus the transmit power $P$ with backhaul capacity $C=6$, coherence time $T=10$, channel gain $\alpha_{ji}=1$ for all $j \in \mathcal{N}_B, i \in \mathcal{N}_M$, and Rayleigh fading channel ($K=0$). It is seen that the performance gains of more complex compression strategies is more evident in the high SNR regime, in which the compression noise imposes a significant bottleneck to the system performance. 

In Fig. \ref{Fig:MultipleBSvsAlpha}, the ergodic achievable sum-rate is plotted versus the inter-cell channel gain $\alpha_{ji}$ assumed to be the same for all $i \neq j$, while $\alpha_{jj} = 1$ for $j \in \mathcal{N}_B$, with backhaul capacity $C=6$, power $P=20dB$, coherence time $T=10$ and Rayleigh fading. As it is well known (see, e.g., \cite{SimeoneBookFnT}), at low inter-cell gain, the inter-cell interference is deleterious; instead, when the inter-cell gain is large enough, the central decoder can take advantage of the additional signal paths and the sum-rate increases.

Finally, we show the impact of the Rician factor $K$ in Fig. \ref{Fig:MultipleBSvsRF} with backhaul capacity $C=6$, power $P=20dB$ and channel gain $\alpha_{ji}=1$ for all $j \in \mathcal{N}_B, i \in \mathcal{N}_M$. We observe that the performance of the joint adaptive compression method approaches that of the joint compression method as the Rician factor $K$ increases. This is because the joint adaptive compression scheme is based on an optimization of the compression strategy that adapts the quantization error on the data signal to the channel estimates for each coherence block. Therefore, in the presence of reduced channel variations due to a larger Rician factor $K$, the performance gain of the adaptive joint approach are reduced. 
\section{Conclusion} \label{Sec:Conclusion}
In this paper, we have studied the design of the backhaul compression strategies for the uplink of network MIMO systems by accounting for both CSI and data transfer from the BSs to the CU. Motivated by the information-theoretic optimization of separate estimation and compression, we have adopted an Estimate-Compress-Forward (ECF) approach, whereby the BSs first estimate the CSI and then forward the compressed CSI to the CU. The alternative Compress-Forward-Estimate (CFE) approach, already studied in previous work, is also considered for reference along with non-coherent transmission. Various schemes of increasing complexity are proposed that aim at optimizing the ergodic achievable sum-rate subject to backhaul constraints. Specifically, separate and joint data signal and CSI compression strategies are devised. Moreover, in the presence of multiple BSs, we have combined the proposed backhaul strategies with distributed source coding to leverage the received signal correlation across BSs. From numerical results, we have observed that the ECF approach outperforms the CFE approach, and that more complex joint compression strategies have significant advantages in the regime of intermediate backhaul capacity, in which the backhaul capacity should be used efficiently, and for sufficiently large SNR and channel coherence times. Finally, we have proposed a semi-coherent strategy that does not convey any CSI or pilot information over the backhaul links. It was seen by numerical results that this scheme is large enough, while the latter is advantageous in the regime of low backhaul capacity.
\appendices
\section{}\label{Appendix:ProofBC;JC}
In this Appendix, we derive equality (\ref{BC_MI;JC_IC}) and the condition $C = C_p + C_d$, with $C_p$ in (\ref{CBC;SC_IC}) and $C_d$ in (\ref{BC;JC_IC}). We start by evaluating $\frac{1}{T} I({\bf{Y}}_{d},\widetilde {\bf{H}};\widehat {\bf{Y}}_{d},\widehat {\bf{H}})$ in (\ref{BC_MI;JC_IC}) as follows:
\begin{eqnarray}
\nonumber \frac{1}{T} I({\bf{Y}}_{d},\widetilde {\bf{H}};\widehat {\bf{Y}}_{d},\widehat {\bf{H}})  &=& \frac{1}{T} \left( I({\bf{Y}}_{d},\widetilde {\bf{H}};\widehat {\bf{H}}) + I({\bf{Y}}_{d},\widetilde {\bf{H}};\widehat {\bf{Y}}_{d}|\widehat {\bf{H}}) \right) \\
\nonumber &=& \frac{1}{T} \left( I(\widetilde {\bf{H}}; \widehat {\bf{H}}) + I({\bf{Y}}_{d};\widehat {\bf{H}}|\widetilde {\bf{H}}) + I({\bf{Y}}_{d},\widetilde {\bf{H}}; \widehat {\bf{Y}}_{d}|\widehat {\bf{H}}) \right) \\
\nonumber &\mathop = \limits^{(a)}& \frac{1}{T} \left( I(\widetilde {\bf{H}}; \widehat {\bf{H}}) + I({\bf{Y}}_{d},\widetilde {\bf{H}}; \widehat {\bf{Y}}_{d}|\widehat {\bf{H}}) \right) \\
\nonumber &=& \frac{1}{T} \left( I(\widetilde {\bf{H}}; \widehat {\bf{H}}) + I({\bf{Y}}_{d}; \widehat {\bf{Y}}_{d}|\widehat {\bf{H}}) + I(\widetilde {\bf{H}}; \widehat {\bf{Y}}_{d}|\widehat {\bf{H}}, {\bf{Y}}_{d}) \right) \\
\nonumber &\mathop = \limits^{(b)}& \frac{1}{T} \left( I(\widetilde {\bf{H}}; \widehat {\bf{H}}) + I({\bf{Y}}_{d}; \widehat {\bf{Y}}_{d}|\widehat {\bf{H}}) \right) \\
&=&\frac{1}{T} \left( I(\widetilde {\bf{H}}; \widehat {\bf{H}}) + h(\widehat {\bf{Y}}_{d}|\widehat {\bf{H}}) - h({\bf{Q}}_{d}) \right), \label{MI_BC;JC_IC}
\end{eqnarray}
where $(a)$ is from the fact that $I({\bf{Y}}_{d};\widehat {\bf{H}}|\widetilde {\bf{H}}) = 0$ due to (\ref{ECS+EE})-(\ref{CCS+QN}), and $(b)$ is form the fact that $I(\widetilde {\bf{H}};\widehat {\bf{Y}}_{d}|\widehat {\bf{H}}, {\bf{Y}}_{d}) = I({\bf{Q}}_{p};{\bf{Q}}_{d}|\widehat {\bf{H}}, {\bf{Y}}_{d}) = I({\bf{Q}}_{p};{\bf{Q}}_{d}) = 0$. Note that $(b)$ proves (\ref{BC_MI;JC_IC}). We can now bound
\begin{eqnarray}
\nonumber h(\widehat {\bf{Y}}_{d}|\widehat {\bf{H}}) &\le& T_d E \left[ \log_2 \left(2 \pi e \right)^{N_{r}} + \log_2 \det \left( \frac{P_d}{N_t} \widehat {\bf{H}} \widehat {\bf{H}}^\dagger + \left( \sigma_{pe}^2 + \sigma_{d}^2 \right) {\bf{I}}_{N_r} \right)\right] \\
&=& T_d \left( N_{r} \log_2 \left( 2 \pi e \right ) + N_{r} \log_2 \left(\sigma_{pe}^2 + \sigma_{d}^2 \right) + E \left [ \log_2 \det \left( {\bf{I}}_{N_r} + \rho_{\textit{eff}} \widehat {\bf{H}} \widehat {\bf{H}}^\dagger \right)\right]  \right), \label{ENT;JC_IC}
\end{eqnarray}
where $\rho_\textrm{eff}$ is defined in (\ref{ESNR;SC_IC}). The inequality in (\ref{ENT;JC_IC}) follows from the maximum entropy theorem because $\widehat {\bf{Y}}_d$ is not Gaussian distributed. Using (\ref{ENT;JC_IC}) in (\ref{MI_BC;JC_IC}) proves the condition $C = C_p + C_d$, with $C_p$ in (\ref{CBC;SC_IC}) and $C_d$ in (\ref{BC;JC_IC}).
\section{}
\label{P_S;JAC_IC} In this Appendix, we solve the non-convex optimization problem of maximizing  (\ref{ESR;JAC_IC}) with respect to ${\bf{R}}_{d}(\widehat {\bf{H}})$ under the constraint $C_p + C_d = C$, with $C_p$ in (\ref{CBC;SC_IC}) and $C_d$ in (\ref{BC;JAC_IC}). We observe that, if $\widehat {\bf{H}}$ was deterministic, the problem would coincide with that solved in \cite[Theorem 1]{Coso09TWC}. The extension to the set-up at hand is then fairly straightforward and is discussed below for completeness.

Following \cite{Coso09TWC}, we first restate the problem in terms of the matrix ${\bf{A}}_{d}(\widehat {\bf{H}})$ defined as ${\bf{R}}_{d}(\widehat {\bf{H}}) = {\bf{A}}_{d}^{-1}(\widehat {\bf{H}}_i)$. By the above definition, the objective function (\ref{ESR;JAC_IC}) is 
\begin{eqnarray}
\nonumber && E \left [ \log_2 \det \left( {\bf{I}}_{N_t} + \frac{P_d}{N_t} \widehat {\bf{H}}^\dagger \left( {\bf{A}}_{d}^{-1}(\widehat {\bf{H}})  + \sigma_{pe}^2 {\bf{I}}_{N_{r}} \right)^{-1} \widehat {\bf{H}} \right) \right] \\
 &=& E \left [ \log_2 \det \left( {\bf{I}}_{N_t} + {\bf{A}}_{d,i}(\widehat {\bf{H}}) \left(\frac{P_d}{N_t} \widehat {\bf{H}} \widehat {\bf{H}}^\dagger + \sigma_{pe}^2 {\bf{I}}_{N_r} \right) \right)  \right]- E \left[ \log_2 \det \left( {\bf{I}}_{N_r} + \sigma_{pe}^2 {\bf{A}}_d(\widehat {\bf{H}}) \right) \right],
\end{eqnarray}
where $\sigma_{pe}^2$ is defined in (\ref{Noise_pe}). The Lagrangian for the problem at hand is hence given as
\begin{eqnarray}
\label{Lagrangian;SBC}\nonumber \mathcal{L} \left( {\bf{A}}_d(\widehat {\bf{H}}), \mu, {\bf{\Upsilon}}(\widehat {\bf{H}}) \right) &=& (1-\mu) E \left [ \log_2 \det \left( {\bf{I}}_{N_r} + {\bf{A}}_d(\widehat {\bf{H}}) \left (\frac{P_d}{N_t} \widehat {\bf{H}} \widehat {\bf{H}}^\dagger + \sigma_{pe}^2 {\bf{I}}_{N_r} \right) \right) \right] \\
&&- E \left[ \log_2 \det \left( {\bf{I}}_{N_r} + \sigma_{pe}^2 {\bf{A}}_d(\widehat {\bf{H}}) \right) \right] + E \left [ \textrm{tr} \left\{ {\bf{\Upsilon(\widehat {\bf{H}})}} {\bf{A}}_d(\widehat {\bf{H}}) \right\} \right],
\end{eqnarray}
with Lagrange multipliers $\mu \ge 0$ for the constraint $C_p + C_d = C$, with $C_p$ in (\ref{CBC;SC_IC}) and $C_d$ in (\ref{BC;JAC_IC}), and ${\bf{\Upsilon}}(\widehat {\bf{H}}) \succeq 0$ for the semidefinite positiveness constraint on ${\bf{A}}_d(\widehat {\bf{H}})$. 

Since the constraint $C_p + C_d = C$, with $C_p$ in (\ref{CBC;SC_IC}) and $C_d$ in (\ref{BC;JAC_IC}), does not define a convex feasible set, the Karush-Kuhn-Tucker (KKT) conditions are only necessary for optimality. In order to solve the problem, therefore, as in \cite{Coso09TWC}, we first find the solution which satisfies the KKT conditions and then show that the derived solution (\ref{EV;JAC_IC}) also satisfies the general sufficiency condition in \cite{BertsekasBook}. Using (\ref{EVD_CM;JAC_IC}), the KKT conditions for the problem at hand can be expressed as 
\begin{subequations}
\label{KKTcondition;SBC}
\begin{eqnarray}
\label{KKT;Stationarity}&& \hspace{-1cm} \left [ \frac{\partial \mathcal{L} \left( {\bf{A}}_d(\widehat {\bf{H}}), \mu, {\bf{\Upsilon}}(\widehat {\bf{H}}) \right) }{\partial {\bf{A}}_d(\widehat {\bf{H}})} \right] = {\bf{0}} \Leftrightarrow \frac{(1-\mu) t_n(\widehat {\bf{H}})}{1 + \lambda_n(\widehat {\bf{H}}) t_n(\widehat {\bf{H}})} - \frac{\sigma_{pe}^2}{1 + \lambda_n(\widehat {\bf{H}}) \sigma_{pe}^2} - \upsilon_n(\widehat {\bf{H}}) = 0, \,\, n = 1, \dots, N_r, \\
&& \hspace{-1cm} \mu \left( E \left [ \log_2 \det \left( {\bf{I}}_{N_r} + {\bf{A}}_d(\widehat {\bf{H}}) \left (\frac{P_d}{N_t} \widehat {\bf{H}} \widehat {\bf{H}}^\dagger + \sigma_{pe}^2 {\bf{I}}_{N_r} \right) \right) \right] - \widetilde C \right) = 0, \\
\label{KKT;Complementary slackness}&& \hspace{-1cm} E \left [ \textrm{tr} \left\{ {\bf{\Upsilon}}(\widehat {\bf{H}}) {\bf{A}}_d(\widehat {\bf{H}}) \right\} \right] = 0 \,\,\,\,\, \Leftrightarrow \,\,\,\,\, \upsilon_n(\widehat {\bf{H}}) \lambda_n(\widehat {\bf{H}}) = 0, \,\,\,\,\, n = 1, \dots, N_r, \\
&& \hspace{-1cm} E \left [ \log_2 \det \left( {\bf{I}}_{N_r} + {\bf{A}}_d(\widehat {\bf{H}}) \left (\frac{P_d}{N_t} \widehat {\bf{H}} \widehat {\bf{H}}^\dagger + \sigma_{pe}^2 {\bf{I}}_{N_r} \right) \right) \right] - \widetilde C \le 0, 
\end{eqnarray}
\end{subequations}
along with $\mu \ge 0$ and ${\bf{\Upsilon}}(\widehat {\bf{H}}) \succeq 0$, where $\widetilde C = \frac{T}{T_d}  ( C - \frac{N_r}{T} \log_2 ( \frac{ \prod_{i=1}^{N_M} ( \sigma_{\widetilde h_{i}}^2)^{N_{t,i}}}{(\sigma_{p}^2)^{N_t}}) )$, we have used the eigendecomposition ${\bf{\Upsilon}}(\widehat {\bf{H}}) = {\bf{U}}(\widehat {\bf{H}}) \textrm{diag} ( \upsilon_1(\widehat {\bf{H}}), \dots,$ $\upsilon_{N_r}(\widehat {\bf{H}}) ) {\bf{U}}^\dagger(\widehat {\bf{H}})$ and we recall that $\lambda_j(\widehat {\bf{H}})$ are the eigenvalues of ${\bf{R}}_d(\widehat {\bf{H}})$. It can be directly shown that the eigenvalues $\lambda_1^*(\widehat {\bf{H}}), \dots, \lambda_{N_r}^*(\widehat {\bf{H}})$ in (\ref{EVD_CM;JAC_IC})-(\ref{EV;JAC_IC}) satisfy the KKT conditions (\ref{KKTcondition;SBC}), if the Lagrange multiplier $\mu^*$ is such that the equality $E \left [ \log_2 \det \left( {\bf{I}}_{N_r} + {\bf{A}}_d(\widehat {\bf{H}}) \left (\frac{P_d}{N_t} \widehat {\bf{H}} \widehat {\bf{H}}^\dagger + \sigma_{pe}^2 {\bf{I}}_{N_r} \right) \right) \right] = \widetilde C$ holds and the Lagrange multipliers $\upsilon_i^*(\widehat {\bf{H}})$ are computed from (\ref{KKT;Stationarity}) and (\ref{KKT;Complementary slackness}). We now show that the derived solution (\ref{EV;JAC_IC}) satisfies also the general sufficiency condition in \cite{BertsekasBook} for optimality. 

\lemma The solution $\left( {\bf{A}}_d^*(\widehat {\bf{H}}) = {\bf{U}}(\widehat {\bf{H}}) \textrm{diag} \left( \lambda_1^*(\widehat {\bf{H}}), \dots, \lambda_{N_r}^*(\widehat {\bf{H}}) \right) {\bf{U}}^\dagger(\widehat {\bf{H}}), \mu^* \right)$ in Proposition \ref{S;JAC_IC} satisfies the sufficiency optimality conditions \cite{BertsekasBook}:
\begin{subequations}
\begin{eqnarray}
&& {\bf{A}}_d^*(\widehat {\bf{H}}) = \arg \max_{{\bf{A}}_d^*(\widehat {\bf{H}}) \succeq 0 } \mathcal{L} \left( {\bf{A}}_d(\widehat {\bf{H}}), \mu^* \right), \\
\label{Suf:KKT_ComSlackness}&& s.t. \,\,\, \mu^* \left( E \left [ \log_2 \det \left( {\bf{I}}_{N_r} + {\bf{A}}_d^*(\widehat {\bf{H}}) \left (\frac{P_d}{N_t} \widehat {\bf{H}} \widehat {\bf{H}}^\dagger + \sigma_{pe}^2 {\bf{I}}_{N_r} \right) \right) \right] - \widetilde C \right) = 0, \\
&& \,\,\,\,\,\,\,\,\,\,\, \mu^* \ge 0,
\end{eqnarray}
\end{subequations}
with the Lagrangian defined as 
\begin{equation}
\mathcal{L} \left( {\bf{A}}_d(\widehat {\bf{H}}), \mu \right) = (1-\mu) E \left [ \log_2 \det \left( {\bf{I}}_{N_r} + {\bf{A}}_d(\widehat {\bf{H}}) \left (\frac{P_d}{N_t} \widehat {\bf{H}} \widehat {\bf{H}}^\dagger + \sigma_{pe}^2 {\bf{I}}_{N_r} \right) \right) \right] - E \left[ \log_2 \det \left( {\bf{I}}_{N_r} + \sigma_{pe}^2 {\bf{A}}_d(\widehat {\bf{H}}) \right) \right]. \label{LagFunc}
\end{equation}
\begin{IEEEproof}
We have inequality $\log \det \left( {\bf{I}} + {\bf{A}} {\bf{B}} \right) \le \log \det \left( {\bf{I}} + {\bf{\Lambda}}_{\bf{A}} {\bf{\Lambda}}_{\bf{B}} \right)$ where ${\bf{A}}, {\bf{B}} \succeq 0$ and ${\bf{\Lambda}}_{\bf{A}}, {\bf{\Lambda}}_{\bf{B}}$ are diagonal matrices with the ordered eigenvalues of ${\bf{A}}$ and ${\bf{B}}$, respectively \cite{Coso09TWC}. As a result, the Lagrangian (\ref{LagFunc}) can be bounded as 
\begin{eqnarray}
\max_{{\bf{A}}_d^*(\widehat {\bf{H}}) \succeq 0 } \mathcal{L} \left( {\bf{A}}_d(\widehat {\bf{H}}), \mu^* \right) \le (1-\mu^*) E \left [ \sum_{i=1}^{N_r}{\log_2 \left( 1 + \lambda_i(\widehat {\bf{H}}) t_i(\widehat {\bf{H}}) \right) } \right] - E \left[ \sum_{i=1}^{N_r}{\log_2 \left( 1 + \lambda_i(\widehat {\bf{H}}) \sigma_{pe}^2 \right)} \right] + \mu^* \widetilde C. \label{maxLagFunc}
\end{eqnarray}
Using this bound and following the same steps as in \cite{Coso09TWC}, we can prove that
\begin{eqnarray}
\mathcal{L} \left( {\bf{A}}_d^*(\widehat {\bf{H}}), \mu^* \right) = (1-\mu^*) E \left [ \sum_{i=1}^{N_r}{\log_2 \left( 1 + \lambda_i^*(\widehat {\bf{H}}) t_i(\widehat {\bf{H}}) \right) } \right] - \left[ \sum_{i=1}^{N_r}{\log_2 \left( 1 + \lambda_i^*(\widehat {\bf{H}}) \sigma_{pe}^2 \right) } \right] + \mu^* \widetilde C \label{optLagFunc}
\end{eqnarray}
for $\left( {\bf{A}}_d^*(\widehat {\bf{H}}), \mu^* \right)$ in Proposition \ref{S;JAC_IC}. It is hence demonstrated that ${\bf{A}}_d^*(\widehat {\bf{H}}) = \arg \max_{{\bf{A}}_d^*(\widehat {\bf{H}}) \succeq 0 } \mathcal{L} \left( {\bf{A}}_d(\widehat {\bf{H}}), \mu^* \right)$ by (\ref{maxLagFunc}) and (\ref{optLagFunc}). Moreover, (\ref{Suf:KKT_ComSlackness}) follows from the condition $C_p + C_d = C$, with $C_p$ in (\ref{CBC;SC_IC}) and $C_d$ in (\ref{BC;JAC_IC}), which concludes the proof.
\end{IEEEproof}
\bibliographystyle{IEEEtran}
\bibliography{refKJK}

\end{document}